\documentclass[a4paper, amsfonts, amssymb, amsmath, reprint, showkeys, nofootinbib, twoside,notitlepage,onecolumn]{revtex4-2}

\usepackage{amsmath,amstext}
\usepackage[T1]{fontenc}
\usepackage{amssymb}
\input{epsf}
\usepackage{graphicx}
\usepackage{ae,aecompl}

\usepackage{hyperref}
\usepackage{amsmath}
\usepackage{amssymb}
\usepackage{mathtools}
\usepackage{bm}
\usepackage{cleveref}
\usepackage{tensor}
\usepackage{braket}
\usepackage{enumitem}
\usepackage{mhchem}
\usepackage{cancel}
\usepackage{amsthm}
\usepackage{upgreek}
\usepackage{mathrsfs}
\usepackage{slashed}

\usepackage{color}

\newcommand{\spc}{\quad \quad \quad}

\newcommand{\Tt}{{\Tilde{t}}}
\newcommand{\Tx}{{\Tilde{x}}}

\def\be{\begin{equation}}
\def\ee{\end{equation}}
\def\beq{\begin{eqnarray}}
\def\eeq{\end{eqnarray}}

\theoremstyle{definition}

\theoremstyle{theorem}

\theoremstyle{corollary}

\begin{document}

\title{Non-covariant parabolic theories of relativistic diffusion}
\author{ L.~Gavassino}
\affiliation{
Department of Mathematics, Vanderbilt University, Nashville, TN, USA
}

\begin{abstract}
A new first-order theory of relativistic dissipation has been recently proposed, where viscous effects are incorporated using the traditional Navier-Stokes framework. Its main novelty is the avoidance of dynamical instabilities by allowing different observers to use equations that are not related by exact Lorentz transformations.  In this work, we explore the implications of this non-covariance in depth. In particular, we discuss how predictions differ between observers moving at nearly luminal speeds relative to each other. We find that all disagreements stem from the relativity of simultaneity, which introduces frame-dependent anisotropic delays in the diffusive process. These anisotropies significantly limit the applicability of the equation used by observers who move very fast relative to the medium. However, the magnitude of the related error remains finite at infinite Lorentz factors, meaning that it is possible to find a regime where all observers agree on the outcome of experiments.
\end{abstract}

\maketitle

\vspace{-0.6cm}
\section{Introduction}
\vspace{-0.4cm}

It is well-known that theories of dissipation whose equations of motion are parabolic, like the diffusion equation
\vspace{-0.1cm}
\begin{equation}\label{diffundo}
\vspace{-0.1cm}
\partial_t \phi =D \partial_x^2 \phi
\end{equation}
(or Navier-Stokes hydrodynamics), are not ``covariantly stable'' in relativity \cite{Hiscock_Insatibility_first_order,Kost2000,GavassinoLyapunov_2020,GavassinoBounds2023}. This just means that, if we transform them under a Lorentz boost, the resulting boosted equations admit unphysical solutions that grow in time. In the specific case of the diffusion equation, it is easy to see this pathology directly. Take \eqref{diffundo}, and apply the transformation $\partial_t = \gamma(\partial_\Tt+v\partial_\Tx)$ and $\partial_x = \gamma(\partial_\Tx+v\partial_\Tt)$, where $v$ is the boost velocity, and $\gamma$ is the Lorentz factor. The result is
\vspace{-0.1cm}
\begin{equation}\label{diffboost}
(\partial_\Tt +v \partial_\Tx)\phi =\gamma D (\partial_\Tx^2+2v\partial_\Tt \partial_\Tx +v^2 \partial_\Tt^2 )\phi \, ,  
\end{equation}
which admits, among its solutions, the runaway profile  $\phi(t)=e^{\Tt/(\gamma v^2 D)}$, signaling an instability \cite{Hiscock_Insatibility_first_order}\footnote{The claim that equation \eqref{diffboost} is unstable simply means that, fixed some generic initial data for $\{\phi,\partial_{\Tilde{t}}\phi\}$ at $\Tilde{t}\, {=}\, 0$, the corresponding solution is overwhelmingly likely to blow up \cite{Hiscock_Insatibility_first_order}. Nonetheless, this does not preclude the existence of well-behaved solutions constructed with different initial-data prescriptions. For instance, one can solve equation \eqref{diffundo} with compactly supported data at $t\, {=}\, 0$, and then boost the resulting solution (typically defined only for $t \,{>}\, 0$) to obtain a regular solution of \eqref{diffboost}, which does not present strong blow-ups.}.


The most common way around such inconveniences is to render the equations hyperbolic, by appropriately modifying the diffusive terms. This idea was first suggested by \citet{cattaneo1958} in 1958, and later became the foundational principle of Israel-Stewart hydrodynamics \cite{Israel_Stewart_1979} (and of nearly all theories of relativistic viscous hydrodynamics \cite{Liu1986,Geroch_Lindblom_1991_causal,Baier2008,Denicol2012Boltzmann,Romatschke2017,FlorkowskiReview2018,GavassinoFronntiers2021,GavassinoUniversalityI2023}). The reason why this works is that instabilities of the kind discussed above (i.e. instabilities that depend on the observer) can arise \textit{only if} the equations of motion propagate information faster than light \cite{GavassinoSuperluminal2021}, which is always the case for parabolic equations \cite{Rauch_book}. Therefore, it is enough to add a ``causality restoration term'' to \eqref{diffundo}, e.g.
\vspace{-0.1cm}
\begin{equation}\label{diffundoconC}
\partial_t \phi =D \big(\partial_x^2 \, {-}\, \alpha \, \partial_t^2 \big) \phi  \spc (\text{with }\alpha \geq 1)  \, ,
\end{equation}
which makes the equation hyperbolic and causal \cite{Rauch_book,Wald,CourantHilbert2_book}, while keeping the long-wavelength behavior unchanged \cite{nagy1994behavior}. With this modification, covariant stability is now guaranteed by the stability-causality theorem \cite[Th 2]{GavassinoSuperluminal2021}. Indeed, if we apply the transformation $\partial_t = \gamma(\partial_\Tt+v\partial_\Tx)$ and $\partial_x = \gamma(\partial_\Tx+v\partial_\Tt)$ to \eqref{diffundoconC}, we now obtain
\begin{equation}\label{diffuboostoC}
(\partial_\Tt+v\partial_\Tx) \phi =\gamma D \big[(1{-}\alpha v^2)\partial_\Tx^2+2v(1{-}\alpha)\partial_\Tt \partial_\Tx +(v^2{-}\alpha) \partial_\Tt^2 \big] \phi \, .
\end{equation}
This time, if we look for solutions of the form $\phi(\Tt)$, we find that none of them grows with $\Tt$, if $\alpha\geq 1$.

Recently, a novel approach has been suggested \cite{Armas:2020mpr,Basar:2024qxd,Bhambure:2024gnf,Bhambure:2024axa}, which does away with causality altogether, while still saving stability. The main idea is the following. Consider a solution of \eqref{diffuboostoC}, whose gradients are very small\footnote{When we say that $\phi(\Tt,\Tx)$ has ``small gradients'', what we mean is that its Fourier transform $\phi(\Tilde{\omega},\Tilde{k})$ is peaked in an infinitesimal neighborhood of the point $(\Tilde{\omega},\Tilde{k})=(0,0)$.}. Then, to first approximation (i.e. to first order in derivatives), we have that $(\partial_\Tt +v \partial_\Tx)\phi \approx 0$. This allows us to replace every $\partial_\Tt$ with $-v\partial_\Tx$ on the right side of \eqref{diffuboostoC}, as the price that we pay is of third order in derivatives. The result is
\vspace{-0.1cm}
\begin{equation}\label{DensityFrame}
\boxed{(\partial_\Tt  +v \partial_\Tx)\phi \approx \dfrac{D}{\gamma^3} \partial_\Tx^2\phi  \, ,}
\end{equation}
which is a stable equation, if solved in the reference frame $(\Tt,\Tx)$.
This suggests that, at long wavelengths and late times, all solutions of \eqref{diffuboostoC} should ``lose memory'' of the exact value of $\alpha$, and they should relax to solutions of the stable parabolic equation \eqref{DensityFrame}, which is sensitive only to $D$ \cite{Geroch1995,LindblomRelaxation1996}. But if that is the case, then it is argued in \cite{Basar:2024qxd} that we may avoid the introduction of $\alpha$ altogether, and we may regard the transformation \eqref{diffundo}$\,\rightarrow\,$\eqref{DensityFrame} as the ``proper way'' of boosting a parabolic equation. Indeed, \eqref{diffundo} and \eqref{DensityFrame} are linked by an approximate Lorentz transformation, in the sense that \eqref{diffboost} and \eqref{DensityFrame} agree up to third-order derivative corrections.

\newpage

The only reason for concern in the above approach is that it is not clear how the truncation error in going from \eqref{diffundo} to \eqref{DensityFrame} grows as $v\rightarrow 1$. If that error were to scale like, say, $\gamma$, then there would always be a fast-moving observer to whom \eqref{DensityFrame} does not work (since $\gamma\rightarrow \infty$ at large $v$). The goal of this article is to settle this matter. More precisely, we aim to provide a \textit{quantitative} answer to the following question: Suppose that an observer (say, Alice) is at rest relative to the medium, and solves \eqref{diffundo}, while another observer (say, Bob) moves at velocity $-v$ relative to the medium, and solves \eqref{DensityFrame}. When these two observers exchange notes, to what extent will their predictions agree?

Throughout the article, we work in natural spacetime units, with $c=1$.

\vspace{-0.2cm}
\section{First test: Expansion of a point-like drop of charge}\label{sectionII}
\vspace{-0.2cm}

In this section, we consider a simple physical scenario that allows us to quantitatively compare the predictions of Alice, who uses \eqref{diffundo}, and Bob, who moves with velocity $-v$ and uses \eqref{DensityFrame}.

\vspace{-0.2cm}
\subsection{Setup of the experiment}\label{ab}
\vspace{-0.2cm}

We consider a standard diffusion experiment, where a drop of charge is injected at the event $(t,x)=(0,0)$, in an otherwise uniform medium. To model this process, Alice writes down a diffusion equation with a localized source,
\begin{equation}\label{hgjdfjddj}
\partial_t \phi =D\partial^2_x \phi +\delta(t)\delta(x) \, ,
\end{equation}
whose solution is the retarded Green's function of the diffusion equation, namely
\begin{equation}\label{Aliciaona}
\phi_{\text{Alice}}(t,x)= \Theta(t) \dfrac{e^{-\frac{x^2}{4Dt}}}{\sqrt{4\pi D t}} \, .
\end{equation}
Bob, instead, solves \eqref{DensityFrame}, with a corresponding source term that is obtained by simply boosting the source in \eqref{hgjdfjddj}:
\begin{equation}\label{sourciamo}
(\partial_\Tt  +v \partial_\Tx)\phi = \dfrac{D}{\gamma^3} \partial_\Tx^2\phi + \dfrac{1}{\gamma} \delta(\Tt)\delta(\Tx) \, .
\end{equation}
The resulting Green function can be easily calculated using Fourier analysis, and we have
\begin{equation}\label{bobone}
\phi_{\text{Bob}}(\Tt,\Tx)=\Theta(\Tt) \sqrt{\dfrac{\gamma}{4\pi D \Tt}} \, e^{-\frac{\gamma^3 (\Tx-v\Tt)^2}{4D\Tt}} \, .
\end{equation}
Of course, to compare \eqref{Aliciaona} and \eqref{bobone}, we must express both solutions in the same coordinates. The most convenient option is to express Bob's solution in Alice's coordinates, which gives
\begin{equation}\label{Bobbonone}
\phi_{\text{Bob}}(t,x)= \Theta(t+vx) \dfrac{e^{\frac{-x^2}{4D(t+vx)}}}{\sqrt{4\pi D (t+vx)}} \, .  
\end{equation}
We see that $\phi_{\text{Alice}}$ and $\phi_{\text{Bob}}$ only differ by a relativity of simultaneity transformation $t \rightarrow t+vx$ (note that $\Tt=\gamma t+\gamma v x$). This discrepancy is no surprise, since both \eqref{diffundo} and \eqref{DensityFrame} assume that the diffusive process is instantaneous in the reference frame of the respective observer, and different observers disagree on what ``instantaneous'' means. The real surprise is that no additional factor $\gamma$ enters \eqref{Bobbonone}, which implies that the limit $v\rightarrow 1$ is well-defined (i.e. nothing diverges).

\vspace{-0.2cm}
\subsection{Two useful benchmarks}\label{sc}
\vspace{-0.2cm}

To get a sense of how ``bad'' the disagreement between \eqref{Aliciaona} and \eqref{Bobbonone} is, we will need to contrast that disagreement with the discrepancy between \eqref{Aliciaona} and alternative (more refined) models of diffusion. One such improved model is SuperBurnett theory \cite{MCLennanBurnett1973} (i.e. third-order hydrodynamics), which posits that
\begin{equation}\label{supbur}
\partial_t \phi=D\partial^2_x \phi -\beta D^3 \partial^4_x \phi \, ,
\end{equation}
where $\beta$ is a third-order transport coefficient, whose value depends on the microscopic interactions. For definiteness, we will set $\beta=1$. Solving \eqref{supbur} in the presence of a localised source $\delta(t)\delta(x)$, we obtain the retarded Green function of SuperBurnett theory:
\begin{equation}\label{supremo}
\phi_{\text{Super}}(t,x)= \Theta(t) \int_{\mathbb{R}} \dfrac{dk}{2\pi} e^{ikx-(Dk^2+D^3k^4)t} \, . \\
\end{equation}
The Fourier integral above does not have an analytical solution, but it can be evaluated numerically. 

\newpage
Another alternative to the diffusion equation is Cattaneo's theory \eqref{diffundoconC}, which explicitly implements a causal bound on propagation. Since Cattaneo's equation of motion is of second order in time, there are 4 retarded Green functions. Here, we are interested in the response of the charge density to a charge injection, which gives \cite{Basar:2024qxd} (setting $\alpha=1$)
\begin{equation}\label{GreenCattuz}
\phi_{\text{Cattaneo}}(t,x) = \Theta(t)\Theta(t^2-x^2) \dfrac{e^{-t/(2D)}}{4D} \bigg[\dfrac{t \, I_1(\tfrac{1}{2D} \sqrt{t^2{-}x^2})}{\sqrt{t^2{-}x^2}} + I_0(\tfrac{1}{2D} \sqrt{t^2{-}x^2}) \bigg]+\Theta(t) \dfrac{e^{-t/(2D)}}{4D} [\delta(x{-}t)+\delta(x{+}t)] \, .
\end{equation}
In practice, one can ignore the Dirac-delta terms, since they describe luminal wavepackets that decay away rapidly.

\vspace{-0.2cm}
\subsection{Quantitative comparison}
\vspace{-0.2cm}

\begin{figure}
    \centering
\includegraphics[width=0.48\linewidth]{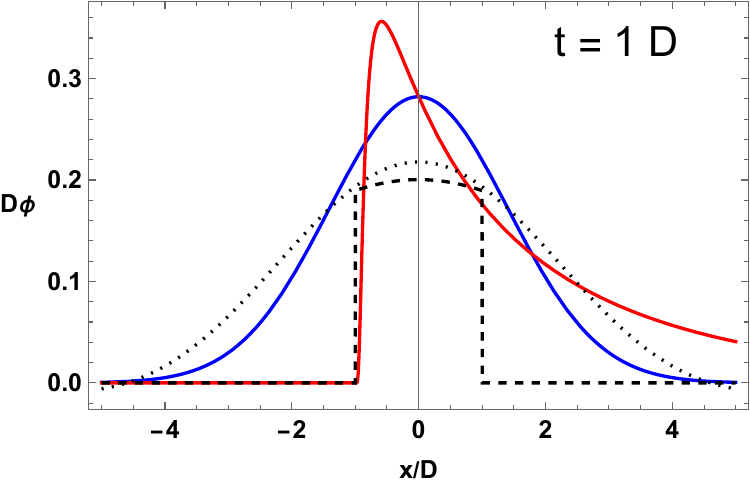}
\includegraphics[width=0.48\linewidth]{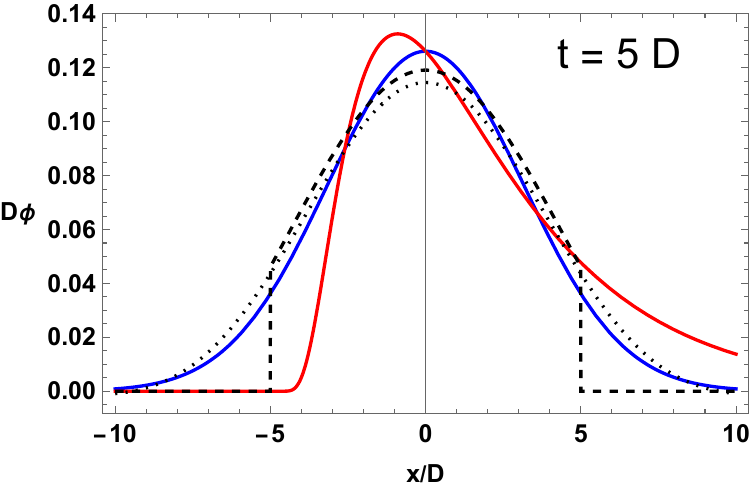}
\includegraphics[width=0.48\linewidth]{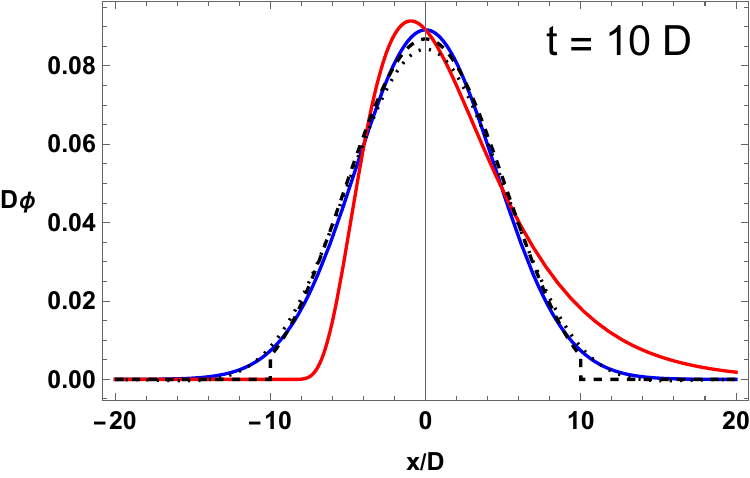}
\includegraphics[width=0.48\linewidth]{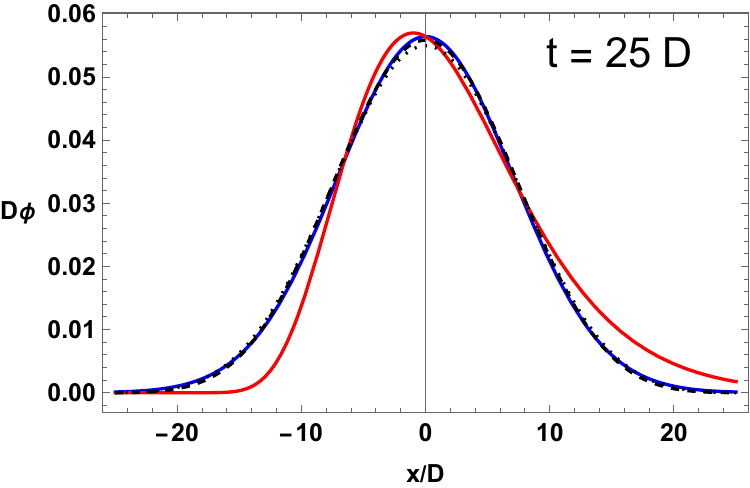}
\includegraphics[width=0.48\linewidth]{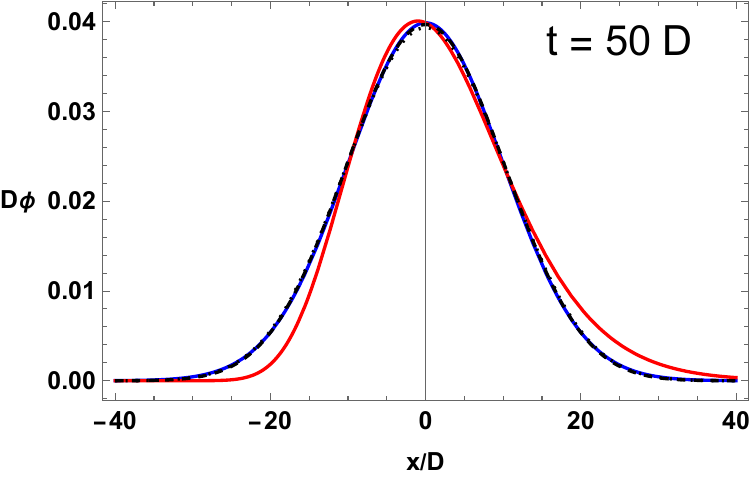}
\includegraphics[width=0.48\linewidth]{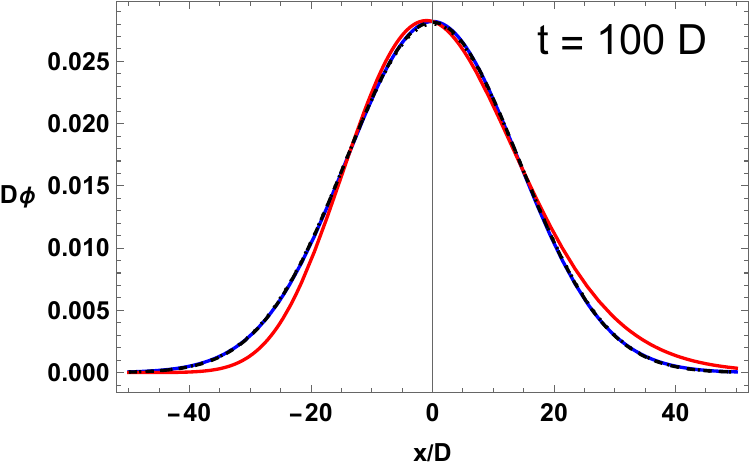}
\includegraphics[width=0.48\linewidth]{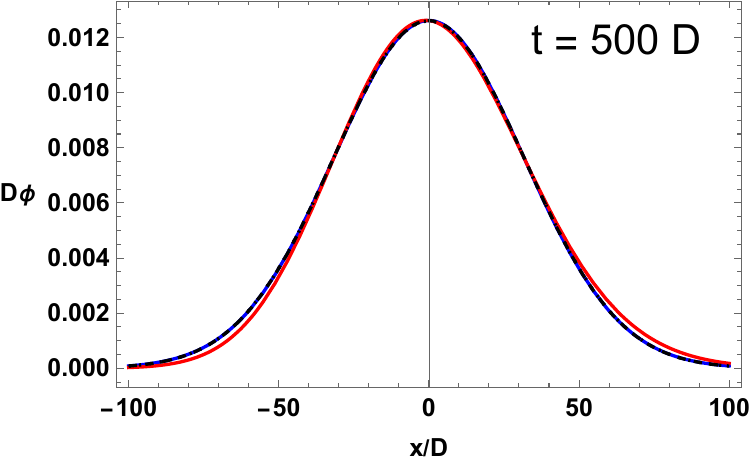}
\includegraphics[width=0.48\linewidth]{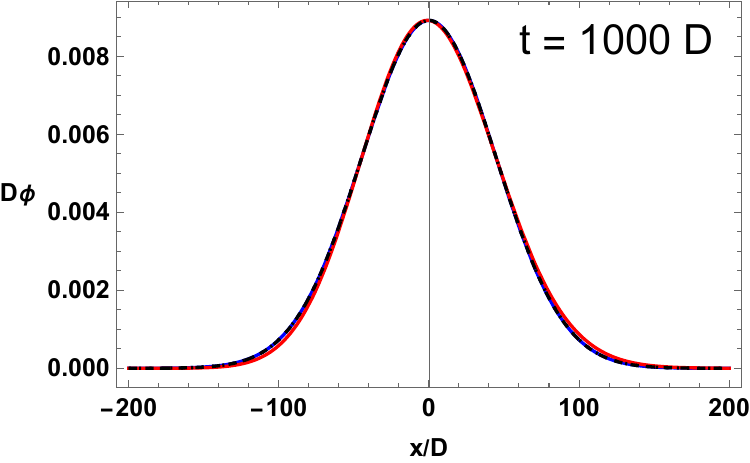}
\caption{Propagation of a drop of charge according to Alice (blue), Bob (red), SuperBurnett (dotted), and Cattaneo (dashed). All snapshots are taken in the reference frame of Alice, who is at rest relative to the medium.}
    \label{fig:greenfunctions}
\end{figure}

Putting together the results of sections \ref{ab} and \ref{sc}, we find ourselves with four alternative predictions for the experimental outcome: $\phi_{\text{Alice}}$, $\phi_{\text{Bob}}$, $\phi_{\text{Super}}$, and $\phi_{\text{Cattaneo}}$. In figure \ref{fig:greenfunctions}, we compare them all, in the limit where Bob's speed $v$ approaches the speed of light (so as to maximize the disagreement between Alice and Bob). We note that, if one introduces the dimensionless coordinates $(T,X)=(t/D,x/D)$ and the rescaled field $\Phi=D\phi$, all dependence on $D$ disappears, and we have a unique scaled behavior for all four theories:
\begin{equation}
\begin{split}
\Phi_{\text{Alice}}(T,X)={}& \Theta(T) \dfrac{e^{-\frac{X^2}{4T}}}{\sqrt{4\pi  T}} \, , \\
\Phi_{\text{Bob}}(T,X)={}& \Theta(T+X) \dfrac{e^{\frac{-X^2}{4(T+X)}}}{\sqrt{4\pi  (T+X)}} \, , \\
\Phi_{\text{Super}}(T,X)={}& \Theta(T) \int_{\mathbb{R}} \dfrac{dK}{2\pi} e^{iKX-(K^2+K^4)T} \, , \\
\Phi_{\text{Cattaneo}}(T,X) ={}& \Theta(T)\Theta(T^2-X^2) \dfrac{e^{-T/2}}{4} \bigg[\dfrac{T\, I_1\big(\tfrac{1}{2} \sqrt{T^2-X^2}\big)}{\sqrt{T^2-X^2}}  + I_0\big(\tfrac{1}{2} \sqrt{T^2-X^2}\big) \bigg] \, .\\
\end{split}
\end{equation}

Let us examine what figure \ref{fig:greenfunctions} reveals. At $t/D = 1$, the predictions of all four models differ significantly. This is unsurprising, as $D$ is usually comparable to the mean free path, so parabolic theories are expected to break down at such scales \cite{Hartman:2017hhp,HellerBounds2023,GavassinoChapmanEnskog2024xwf}. These effects diminish as time progresses, and by $t/D=25$, the predictions of the Alice, SuperBurnett, and Cattaneo models become nearly indistinguishable. Bob's model, however, continues to diverge noticeably from the others. It is not until around $t/D\sim 1000$ that this discrepancy finally disappears. This brings us to the conclusion that truncating the gradient expansion in the local rest frame (as in Alice's theory) yields significantly more accurate predictions than doing so in a highly boosted frame (as in Bob's theory). Nevertheless, there is a region of spacetime ($t\gtrsim 1000D$) where all possible ``Bobs'' agree with Alice, even those with an infinitely high boost rapidity.

\begin{figure}
    \centering
\includegraphics[width=0.55\linewidth]{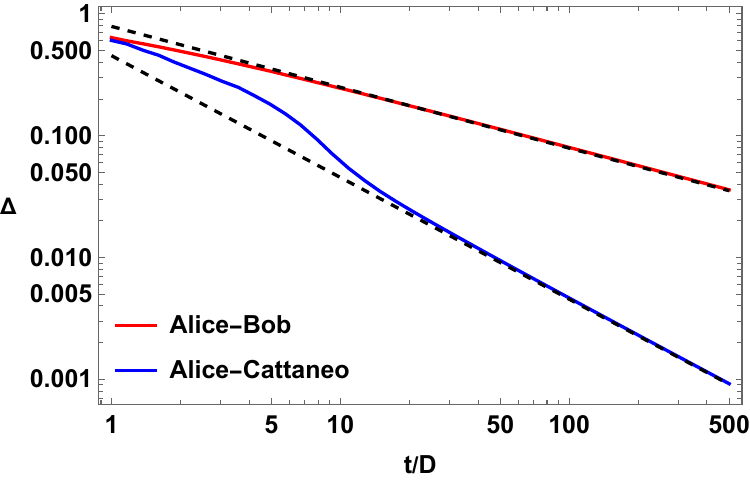}
    \caption{Comparison between the Alice-Bob discrepancy (Red) and the Alice-Cattaneo discrepancy (Blue), as defined in \eqref{discrepiamo}. At late times, the Alice-Bob discrepancy scales like $\text{const}\times \sqrt{D/t}$ (upper dashed), while the Alice-Cattaneo discrepancy scales like $\text{const} \times D/t$ (lower dashed).}
    \label{fig:errorone}
\end{figure}

The above discussion can be made more precise by introducing the ``discrepancy functions''
\begin{equation}\label{discrepiamo}
\begin{split}
\Delta_{\text{Alice,Bob}}(T)={}& \int_{\mathbb{R}} |\Phi_{\text{Alice}}(T,X)-\Phi_{\text{Bob}}(T,X)| \, dX \, , \\
\Delta_{\text{Alice,Cattaneo}}(T)={}& \int_{\mathbb{R}} |\Phi_{\text{Alice}}(T,X)-\Phi_{\text{Cattaneo}}(T,X)| \, dX \, , \\
\end{split}
\end{equation}
which measure the disagreement between Alice, Bob, and Cattaneo quantitatively. As can be seen from figure \ref{fig:errorone}, the discrepancy between Alice and Cattaneo decays like $D/t$, while the disagreement between Alice and Bob decays like $\sqrt{D/t}$, which is much slower. In section \ref{secondonetestone}, we will show that this trend is due to the truncation error of the transformation \eqref{diffundo}$\rightarrow$\eqref{DensityFrame} being of lower order in gradients than the truncation error of the transformation \eqref{supbur}$\rightarrow$\eqref{diffundo}.

\vspace{-0.2cm}
\subsection{From Bob's perspective}\label{fromBobseperro}
\vspace{-0.2cm}

It is important to note that our main result, namely that Alice, SuperBurnett and Cattaneo converge much earlier than Bob, is a Lorentz invariant statement. In fact, if we pick an event in spacetime, e.g. $(t,x)=(25 D,20 D)$, and we find that at such event $\phi_{\text{Alice}}\approx \phi_{\text{SuperBurnett}}\approx \phi_{\text{Cattaneo}}\ll \phi_{\text{Bob}}$, then all observers will agree that, at such event, Bob's model is overestimating $\phi$. The only difference is that different observers will assign different coordinates to that event. For example, if Bob moves with velocity $-v=-0.999$, then the event above becomes $(\Tilde{t},\Tilde{x})\approx (1006 D,1006 D)$. This immediately tells us that, from Bob's perspective, it takes much longer than $\Tilde{t}=1000D$ for his theory to converge to the other theories. Indeed, since most of the support of $\phi$ is localized in a neighbourhood of Alice's worldline $x=0$, the standard formula of time dilation applies, meaning that, from Bob's perspective, the Alice-Cattaneo discrepancy scales like $\gamma D/\Tt$, while the Alice-Bob discepancy scales like $\sqrt{\gamma D/\Tt\, \,}$. Consequently, a highly boosted Bob must wait a considerable amount of time before entering the region where his predictions become meaningful.


\vspace{-0.2cm}
\section{Second test: Comparison in Fourier space}\label{secondonetestone}
\vspace{-0.2cm}

The previous test has revealed that, in the case of an expanding drop of charge, the discrepancy between \eqref{DensityFrame} and \eqref{diffundo} is much larger than the discrepancy between \eqref{diffundo} and higher-order theories of diffusion, like \eqref{diffundoconC} and \eqref{supbur}. However, such discrepancy remains finite as $v\rightarrow 1$. We will now use Fourier analysis to show that this feature is general.

\vspace{-0.2cm}
\subsection{Truncation error estimates}\label{vDkbeta}
\vspace{-0.2cm}

In a generic microscopic theory (like kinetic theory), where the equilibrium state is isotropic in Alice's reference frame, the dispersion relation $\omega(k)$ (as viewed by Alice) is even under the 180$^{\text{o}}$-rotation $k\rightarrow -k$. Hence, when we Taylor expand $\omega(k)$ for small $k$, we obtain 
\begin{equation}
\omega=-iDk^2-i\beta D^3 k^4 +\mathcal{O}(k^6) \, .
\end{equation}
Thus, the relative error that we commit when we truncate $\omega$ at order $k^2$ is
\begin{equation}\label{microAlice}
\frac{\omega_{\text{microscopic}}-\omega_{\text{Alice}}}{\omega_{\text{Alice}}} \sim \beta D^2 k^2 \, .
\end{equation}
Consider now equation \eqref{DensityFrame}, and express it in Alice's coordinates. The result is
\begin{equation}\label{boninAlicetrame}
\partial_t\phi =D(\partial_x{-}v\partial_t)^2 \phi \, ,
\end{equation}
which, not surprisingly, differs from \eqref{diffundo} only by a relativity of simultaneity transformation $\partial_x \rightarrow \partial_x -v \partial_t$. Now, when we compute the dispersion relations of \eqref{boninAlicetrame}, we obtain two functions $\omega(k)$, one of which is gapped and unstable. However, if we restrict our attention to solutions that are finite in Bob's frame, and we look only at the part of these solutions that comes \textit{after} the injection of charge by external sources, the contribution of the gapped modes vanishes in Alice's frame \cite{GavassinoSuperluminal2021,Hiscock_Insatibility_first_order,Kost2000}. Indeed, the Green function \eqref{Bobbonone} contains no gapped contribution for $t\,{>}\,0$ (see Appendix~\ref{appendixAAA}). Therefore, all the truncation error comes from the expansion of the gapless dispersion relation, which reads
\begin{equation}\label{vd2k3}
\omega=-iDk^2-2vD^2 k^3 +\mathcal{O}(k^4) \, .
\end{equation}
From this, we conclude that the relative error that we commit when we truncate $\omega$ at order $k^2$ is
\begin{equation}\label{bobalrespectively}
\frac{\omega_{\text{Bob}}-\omega_{\text{Alice}}}{\omega_{\text{Alice}}} \sim vDk \, ,
\end{equation}
where factors $\pm i$ are irrelevant for estimating the \textit{magnitude} of the error. We see that, at relativistic velocities, the discrepancy between Alice and Bob scales like the square root of the discrepancy between Alice and the microscopic theory. This makes \eqref{DensityFrame} a much less accurate theory than ordinary diffusion. Nevertheless, such discrepancy grows with $v$, and not with $\gamma$, so it remains finite as $v\rightarrow 1$. In other words, there is no exchange of limit problem: 
\begin{equation}
\lim_{k\rightarrow 0} \lim_{v\rightarrow 1}\frac{\omega_{\text{Bob}}-\omega_{\text{Alice}}}{\omega_{\text{Alice}}}=\lim_{v\rightarrow 1}\lim_{k\rightarrow 0} \frac{\omega_{\text{Bob}}-\omega_{\text{Alice}}}{\omega_{\text{Alice}}}=0 \, .
\end{equation}

\subsection{Expansion of the drop of charge revisited}\label{expansionrevisited}
\vspace{-0.2cm}

We are now able to explain the behavior in figure \ref{fig:errorone} from standard Fourier analysis. To this end, consider a generic function $\phi(t,x)$, which is a superposition of gapless modes only:
\begin{equation}
\phi(t,x)=\int \dfrac{dk}{2\pi} \phi(k) e^{ikx-(Dk^2+...)t} \, .
\end{equation}
At late times, the most relevant contributions to the integral come from those modes for which $Dk^2t \lesssim 1$, since all modes with larger $k$ are exponentially suppressed by the factor $e^{-Dk^2 t}$. On the other hand, modes with larger $k$ experience larger truncation error. Thus, most of the truncation error comes from those modes with $k^2 \sim (Dt)^{-1}$. Plugging this choice of $k$ into \eqref{microAlice} and \eqref{bobalrespectively} respectively, we obtain an error that scales like $\beta D/t$ for the discrepancy between Alice and Cattaneo, and like $v\sqrt{D/t}$ for the discrepancy between Alice and Bob. This shows that the behavior observed in figure \ref{fig:errorone} is general, and it extends beyond the limited scenario discussed in section \ref{sectionII}.

\vspace{-0.2cm}
\section{Consistency check: Local equilibration in Bob's frame}
\vspace{-0.2cm}

Before moving to the last test, an interesting issue needs to be addressed. In section \ref{fromBobseperro}, we showed that, if it takes a time $\Delta t$ for Bob's solution to agree with Cattaneo's solution in Alice's frame, then, by time dilation, it will take $\Delta \Tt\sim \gamma \Delta t$ for the same to happen in Bob's frame. This seems to contradict the analysis of \cite{Basar:2024qxd}, where it is argued that, instead, Bob's equation should become valid after a \textit{contracted} time interval $\Delta \Tt\sim \Delta t/\gamma$. In this section, we resolve such apparent contradiction.

\vspace{-0.2cm}
\subsection{Outline of the argument}
\vspace{-0.2cm}

Let us first recap how the authors of \cite{Basar:2024qxd} arrived at the estimate $\Delta\Tt \sim \Delta t/\gamma$. 

Consider the Cattaneo equation \eqref{diffuboostoC}, as viewed in Bob's reference frame. Set $\alpha=1$ for simplicity, so that
\begin{equation}\label{cattaneoalphais1}
(\partial_\Tt+v\partial_\Tx) \phi =\dfrac{D}{\gamma} \big(\partial_\Tx^2- \partial_\Tt^2 \big) \phi \, .
\end{equation}
Then, look for solutions where $\phi$ does not depend on $\Tx$. One such solution is just $\phi=\text{const}$, and the other one is $\phi\propto \exp(-\gamma \Tt/D)$. This tells us that the non-hydrodynamic mode of the boosted Cattaneo equation decays over a contracted timescale $\Delta \Tt =D/\gamma$, suggesting that local equilibration is faster in Bob's frame than it is Alice's frame.

Before addressing this result in detail, we would like point out a minor flaw in the above argument. That is, the ``time contracted'' decay law appears only in the particular case where $\alpha=1$. If we repeat the above calculation with a generic $\alpha>1$, we obtain a non-hydrodynamic mode that decays like $\phi \propto \exp[-\Tt/(\gamma D(\alpha-v^2))]$, which gives us the equilibration timescale $\Delta \Tt=\gamma D(\alpha -v^2)$. As $v\rightarrow 1$, such timescale becomes $\Delta \Tt=\gamma D(\alpha -1)$, which \textit{grows} with $\gamma$, thereby making the time contracted law a very special limit (an interesting limit, nonetheless).

\vspace{-0.2cm}
\subsection{Resolution: Relativity of simultaneity, once again}
\vspace{-0.2cm}

In \cite{Basar:2024qxd}, the authors focus on the non-hydrodynamic mode $\phi\propto \exp(-\gamma \Tt/D)$, as a measure of how the fluid locally equilibrates in Bob's frame. This surely feels natural to Bob, since the given mode has no spatial gradients, so it describes a process of homogeneous relaxation to equilibrium. However, because of relativity of simultaneity, other observers attribute to this mode large gradients \textit{also in space}. For example, in Alice's coordinates, this mode reads
\begin{equation}\label{gringonekrr}
\phi \propto e^{-\gamma^2 (t+vx)/D} \, ,
\end{equation}
whose gradients in space diverge at large $v$. Hence, the mode $\phi\propto \exp(-\gamma \Tt/D)$ describes a rather special scenario, where the system has been finely tuned to have small space-gradients and large time-gradients \textit{only} in Bob's frame. In a more realistic situation, where the non-equilibrium effects have not been ``designed'' for Bob, time and space derivatives should have similar magnitudes in highly boosted frames. Therefore, Bob should also look at those contributions coming from generic modes with large $\partial_\Tx$, as such modes may just be the ``relaxation modes'' of someone else. According to \eqref{cattaneoalphais1}, the longest-lived of such modes decays like
\begin{equation}\label{sbescek}
\phi \propto e^{-\Tt/(4D\gamma)}  \spc (\text{if }\partial_\Tx \rightarrow \infty) \, .
\end{equation}
Thus, if Bob encounters a solution $\phi(\Tt,\Tx)$ that initially contains all modes (as in our drop of section \ref{sectionII}), he has to wait a \textit{dilated} timescale $\Delta \Tt \sim  \gamma D$ for the non-hydrodynamic contributions coming from \eqref{sbescek} to decay away, thereby resolving the apparent contradiction.

\section{Third test: Expansion of a finite drop of charge}

One potential criticism of the test in Section \ref{sectionII} is that the initial development of the shape of $\phi$ takes place outside the hydrodynamic regime (when $t/D \sim 1$), and only becomes hydrodynamic at later times. As a result, the disagreement observed between Alice and Bob at $t/D \sim 100$ could simply be a remnant of the early-time mismatch, rather than evidence that the equations differ at that length-scale. Given the discussion in section \ref{secondonetestone}, it should already be clear that this explanation cannot account for the entirety of the disagreement. Nevertheless, to resolve the issue conclusively, it may be helpful to consider a scenario in which the system remains within the hydrodynamic regime throughout its entire evolution.

\subsection{Setup of the new experiment}

We consider a generalization of the experiment outlined in section \ref{sectionII}, where the drop of charge now has a finite size, and is inserted ``slowly'' into the medium. In this way, the source terms on the right side of \eqref{hgjdfjddj} and \eqref{sourciamo} are smooth, and their Fourier transform is supported at long wavelengths both in space and in time\footnote{In a non-relativistic setting, there would be no need to assume that the drop is injected slowly. However, in a relativistic framework, an instantaneous injection of an extended drop is converted, after a Lorentz boost, into a sequence of instantaneous injections of Dirac-delta drops (by relativity of simultaneity), and this results in large spatial gradients as viewed by Bob.}. The resulting solutions can be expressed as convolutions of the respective Green functions with the new source:
\begin{equation}\label{easonino}
\phi_{\text{Alice/Bob/Cattaneo}}(t,x)=\int_{\mathbb{R}^2}\phi_{\text{Alice/Bob/Cattaneo}}^{\text{point-like drop}}(t{-}t',x{-}x') \mathcal{S}(t',x') dt' dx' \, , 
\end{equation}
where $\phi_{\text{Alice/Bob/ Cattaneo}}^{\text{point-like drop}}$ are the solutions computed in section \ref{sectionII}. For the shape of the source, we choose the compactly supported function $\mathcal{S}(t,x)=\text{Bump}(\sqrt{t^2{+}x^2}/R)$, where the radius $R$ defines the lengthscale of the problem, and
\begin{equation}
\text{Bump}(r)=\begin{cases}
e^{-\frac{1}{1-r^2}} & \text{if }r<1 \, , \\
0 & \text{if }r\geq 1 \, . \\
\end{cases}
\end{equation}
In the following, we will focus on the profile of $\phi$ right after the drop has been injected, namely at $t=R$. It follows that, in the case of Cattaneo, the Dirac-delta wavefronts in \eqref{GreenCattuz} should no longer be discarded, as these may add up to a non-negligible (long-wavelength) contribution to the total density profile at time $R$.


\subsection{Bob has an interesting idea}

In the scenario above, where the source varies slowly, Bob may develop an alternative model that slightly refines the straightforward convolution of the Green function \eqref{Bobbonone} with $\mathcal{S}$, resulting in more accurate predictions. To see how, let us consider Alice's equation of motion with a source,
$\partial_t \phi =D\partial^2_x \phi +\mathcal{S}$, and let us perform an \textit{exact} boost:
\begin{equation}
(\partial_\Tt +v \partial_\Tx)\phi =\gamma D (\partial_\Tx^2+2v\partial_\Tt \partial_\Tx +v^2 \partial_\Tt^2 )\phi+\mathcal{S}/\gamma \, .
\end{equation}
To arrive at his equation, Bob would like to replace $\partial_\Tt\phi$ with $-v\partial_\Tx \phi$ in the second derivative terms. However, this would not be fully consistent in the presence of $\mathcal{S}$. To be consistent, he should instead write $\partial_\Tt \phi=-v\partial_\Tx \phi+\mathcal{S}/\gamma+\mathcal{O}(\partial^2)$. With this new replacement rule, Bob's novel equation also contains modifications to the source term:
\begin{equation}
(\partial_\Tt  +v \partial_\Tx)\phi = \dfrac{D}{\gamma^3} \partial_\Tx^2\phi+\big[1+ D\gamma(2v{-}v^3)\partial_\Tx +D\gamma v^2 \partial_\Tt \big]\dfrac{\mathcal{S}}{\gamma} \, .
\end{equation}
Expressing this equation in Alice's coordinates, we then obtain
\begin{equation}
\partial_t \phi =D(\partial_x{-}v\partial_t)^2\phi +\big[1+Dv(2\partial_x{-}v\partial_t) \big]\mathcal{S} \, ,
\end{equation}
whose solution can be expressed as the following convolution:
\begin{equation}\label{difficultinon}
\phi_{\text{Bob}}(t,x)=\int_{\mathbb{R}^2} \big[1+Dv(2\partial_x{-}v\partial_t) \big]\phi_{\text{Bob}}^{\text{point-like drop}}(t{-}t',x{-}x') \mathcal{S}(t',x') dt' dx' \, ,
\end{equation}
where $\phi_{\text{Bob}}^{\text{point-like drop}}$ is the retarded Green function given in \eqref{Bobbonone}.

\newpage
\subsection{Quantitative comparison}
\vspace{-0.3cm}

\begin{figure}[h!]
    \centering
\includegraphics[width=0.46\linewidth]{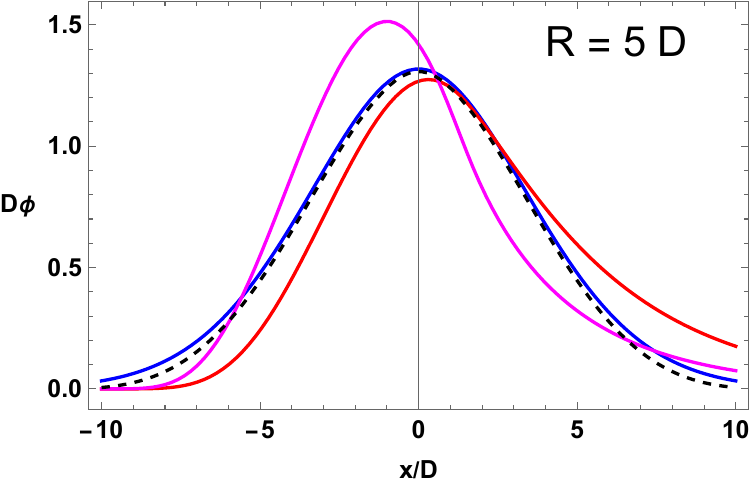}
\includegraphics[width=0.46\linewidth]{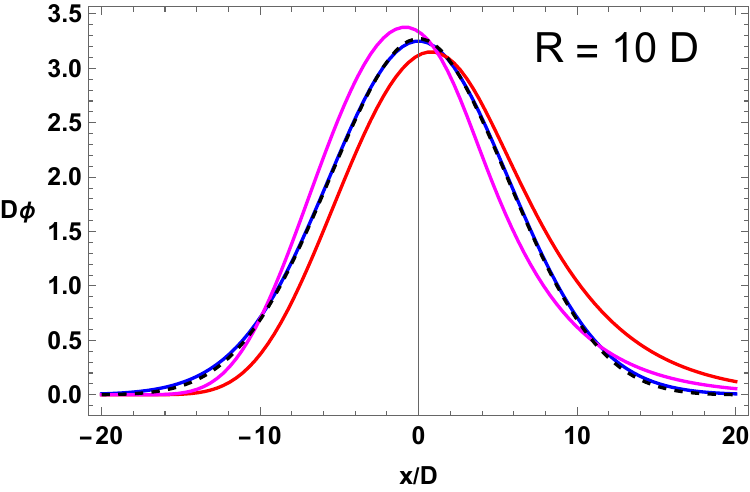}
\includegraphics[width=0.46\linewidth]{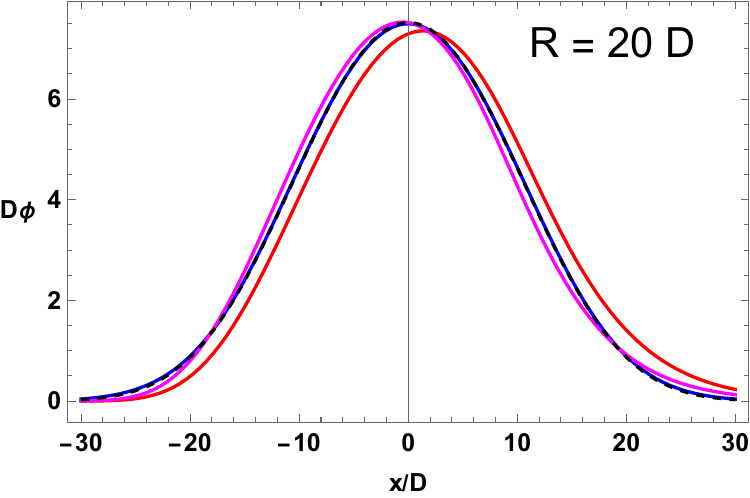}
\includegraphics[width=0.46\linewidth]{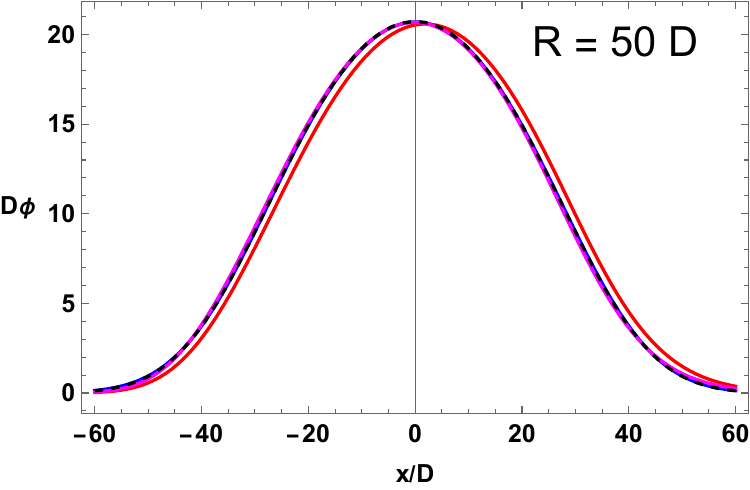}
\includegraphics[width=0.46\linewidth]{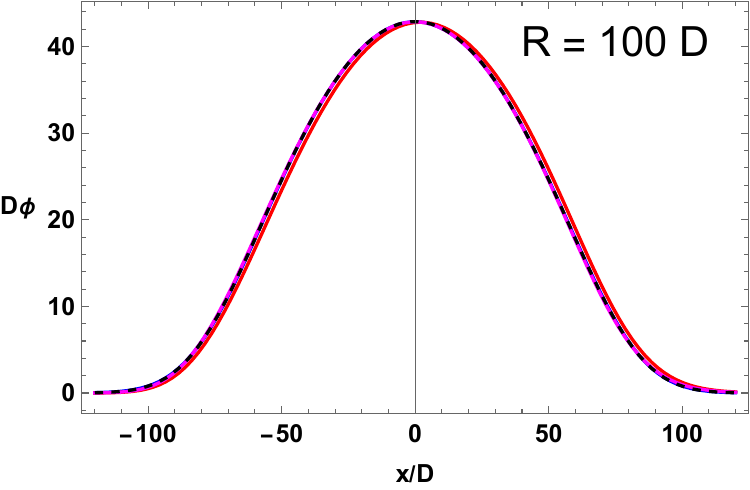}
\includegraphics[width=0.46\linewidth]{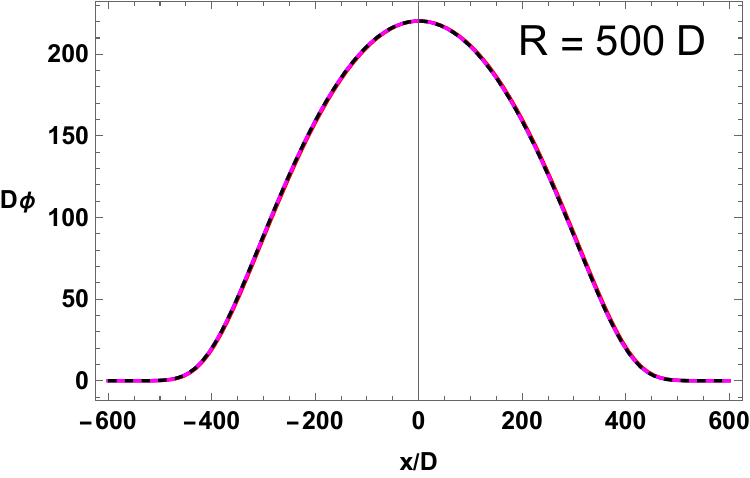}
\caption{Shape of a drop of charge of size $\sim 2R$ which has been slowly injected for a time $2R$. The snapshots show the density profile right after the injection has ended, and they compare the predictions of Alice (blue), Bob who uses \eqref{easonino} (red), Bob who uses \eqref{difficultinon} (magenta), and Cattaneo (dashed). }
    \label{fig:Convoluted}
\end{figure}

The drop shapes at the end of the injection process, as predicted by Alice, Bob (with $v=1$), and Cattaneo, are compared in Figure \ref{fig:Convoluted}. The behaviours are analogous to those in figure \ref{fig:greenfunctions}. Alice agrees with Cattaneo at a lengthscale of about $R/D\sim 10$. Bob, on the other hand, achieves similar accuracies for $R/D \sim 50$ if he uses the ``refined solution'' \eqref{difficultinon}, and for $R/D\sim 100$ if he uses the ``naive convolution'' \eqref{easonino}.

Figure \ref{fig:Convoluted} helps illustrate that the discrepancies observed in figure \ref{fig:greenfunctions} cannot be solely attributed to early-time non-hydrodynamic behavior. Consider, for instance, the snapshot at $t/D=50$ in figure \ref{fig:greenfunctions}. At that time, the characteristic length-scale of the drop is approximately $20 \, D$. Accordingly, we should compare this snapshot with the case $R/D=20$ in figure \ref{fig:Convoluted}. The discrepancies are comparable in both cases: Alice and Cattaneo coincide almost perfectly, while Bob remains noticeably farther away. Remarkably, the refinement \eqref{difficultinon} has significantly improved Bob's prediction. However, this cannot fully compensate for the fact that Bob's error is of lower order in gradients relative to the Alice-Cattaneo discrepancy (as shown in section \ref{vDkbeta}).

\newpage
\section{Sound waves}
\vspace{-0.2cm}

Till this point, we have been focusing on equation \eqref{DensityFrame}, which is a model for diffusive processes as viewed in boosted reference frames. However, hydrodynamic systems also admit another type of hydrodynamic excitation: sound waves. Let us see if any of our main conclusions change when sound waves are considered. 

As before, we aim to compare the predictions of two observers: Alice, who is at rest relative to the medium and solves linearized Navier-Stokes hydrodynamics, and Bob, who moves with velocity $-v$, and solves an ``approximately boosted'' version of linearized Navier-Stokes. Again, the discrepancy between Alice and Bob shall be contrasted with the discrepancy between Alice and a higher-order viscous theory.

\vspace{-0.2cm}
\subsection{From Burnett to Alice}
\vspace{-0.2cm}

In the case of sound waves, we do not need to go to third order in gradients to see modifications to Navier-Stokes. Burnett's theory \cite{Struchtrup2011} (i.e. second-order hydrodynamics) is enough for that purpose. If the sound wave travels to the right, the Burnett linearised equation of motion (expressed in Alice's reference frame) reads
\begin{equation}\label{soundburnett}
(\partial_t+c_s\partial_x)\phi =D\partial^2_x \phi +\beta D^2 \partial^3_x \phi \, ,  
\end{equation}
where $c_s$ is the speed of sound, and $\beta$ is a second-order transport coefficient. Similarly to what we did in section \ref{sectionII}, we imagine that an external force impresses a ``kick'' at the event $(t,x)=(0,0)$, which generates a sound wavepacket in the fluid. As before, we model such kick as a term $\delta(t)\delta(x)$ on the right of \eqref{soundburnett}, which produces the solution
\begin{equation}\label{grinburnettwithsound}
\phi_{\text{Burnett}}(t,x)=\Theta(t) \int \dfrac{dk}{2\pi} e^{ik(x-c_s t-\beta D^2k^2 t)-Dk^2 t} \, .
\end{equation}
This is our ``higher-order prediction''.  

To obtain Alice's prediction, we only need to drop the third-derivative term (i.e. set $\beta=0$) in \eqref{soundburnett}. Then, the Fourier integral in \eqref{grinburnettwithsound} can be evaluated analytically, and we obtain
\begin{equation}
\phi_{\text{Alice}}(t,x)=\Theta(t) \dfrac{e^{-\frac{(x-c_s t)^2}{4Dt}}}{\sqrt{4\pi Dt}} \, .
\end{equation}
It is important to note that, this time, the truncation error in going from the higher-derivative theory to Alice's theory scales like $D^2 k^3$, while for diffusive modes it scaled like $D^3 k^4$ (see discussion in section \ref{secondonetestone}). Therefore, in the case of sound waves, the Alice-Burnett discrepancy will decay like $\sqrt{D/t}$, by the argument of section \ref{expansionrevisited}. This is the same scaling as that of the error in going from Alice to Bob in diffusive modes (again, see section \ref{expansionrevisited}). For this reason, we no longer expect Alice's solution to beat Bob's solution in a formal ``order of magnitude'' hierarchy.

\vspace{-0.2cm}
\subsection{From Alice to Bob}
\vspace{-0.2cm}

\begin{figure}
    \centering
\includegraphics[width=0.48\linewidth]{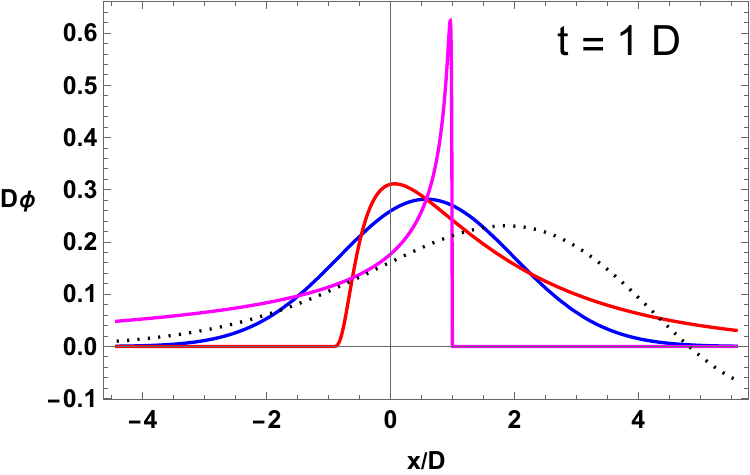}
\includegraphics[width=0.48\linewidth]{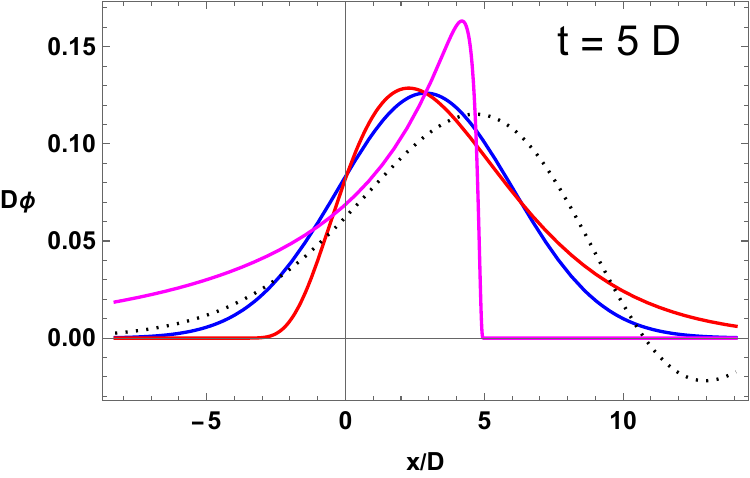}
\includegraphics[width=0.48\linewidth]{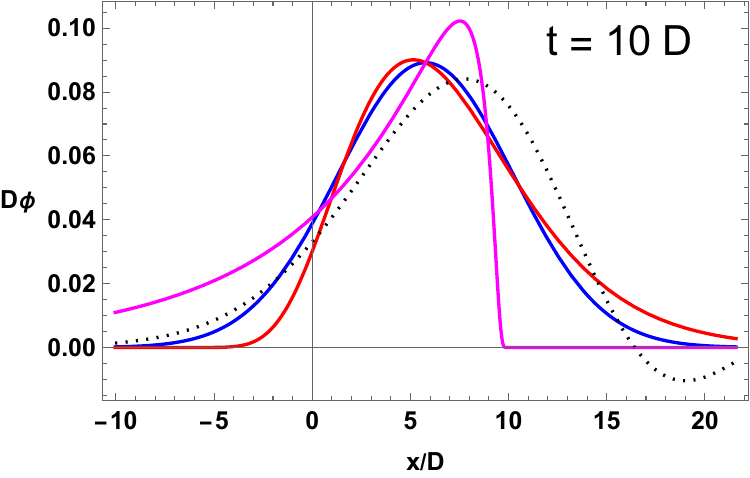}
\includegraphics[width=0.48\linewidth]{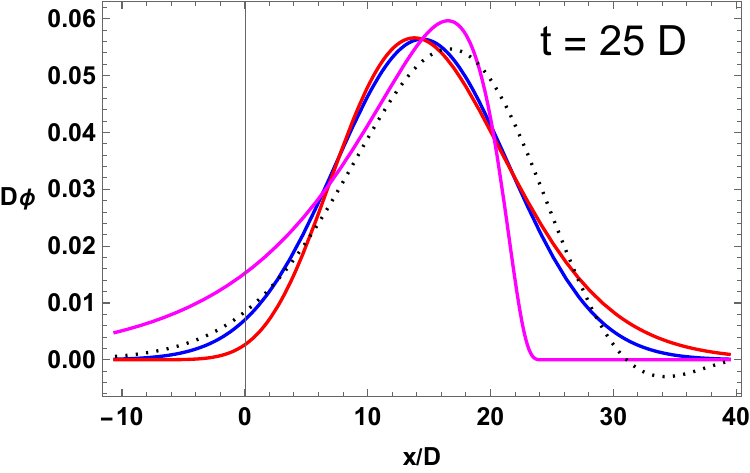}
\includegraphics[width=0.48\linewidth]{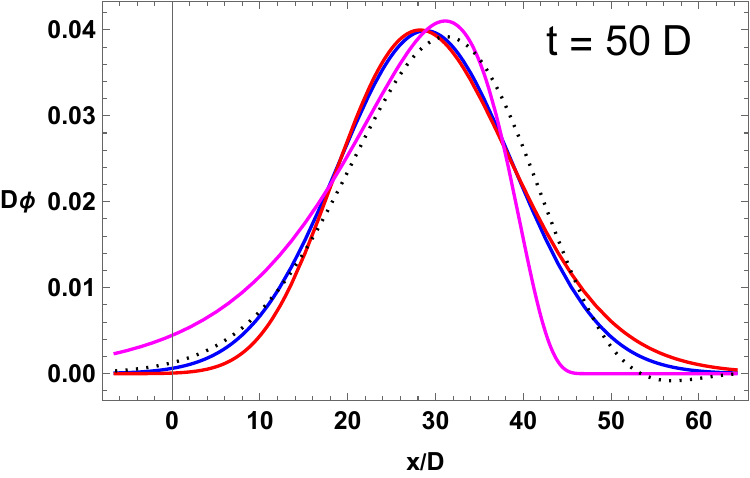}
\includegraphics[width=0.48\linewidth]{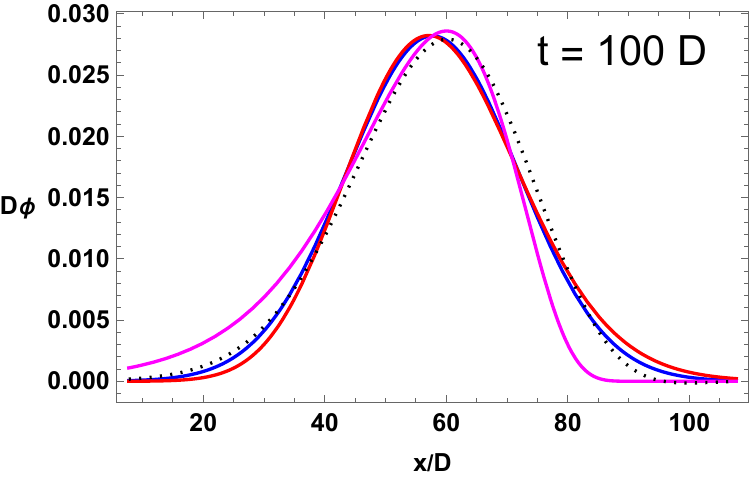}
\includegraphics[width=0.48\linewidth]{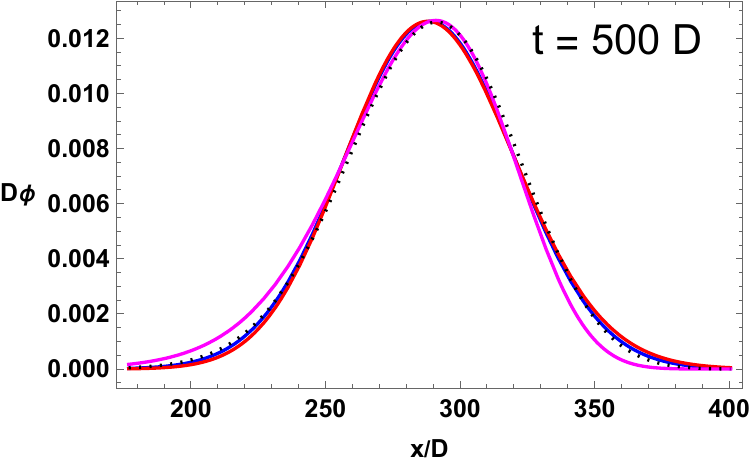}
\includegraphics[width=0.48\linewidth]{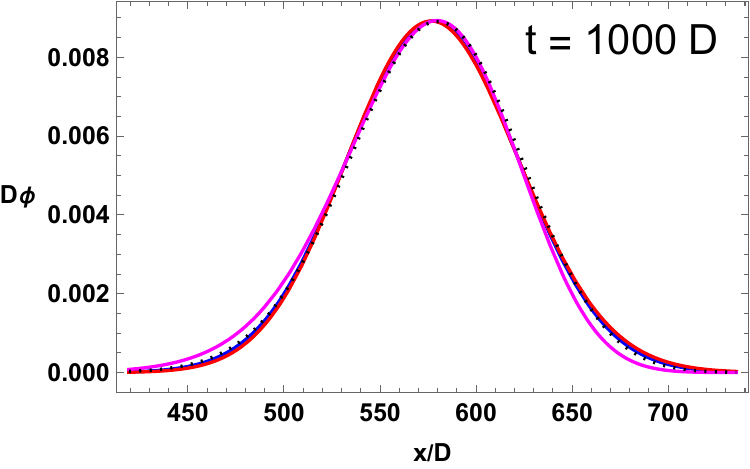}
\caption{Propagation of a right-moving (initially point-like) sound wavepacket according to Alice (blue), Bob with $v=1^-$ (red), Bob with $v=-1^+$ (magenta), and Burnett. All plots are in the reference frame of Alice.}
    \label{fig:soundiamo}
\end{figure}

To derive Bob's theory, we perform an exact boost on equation \eqref{soundburnett} (assuming $\beta=0$),
\begin{equation}
\bigg(\partial_\Tt +\dfrac{c_s+v}{1+c_s v}\partial_\Tx \bigg)\phi =\dfrac{\gamma D}{1+c_s v} (\partial_\Tx +v \partial_\Tt)^2 \phi \, ,
\end{equation}
and we use the first-order equation of motion to replace the time derivative on the right. The result is
\begin{equation}\label{garrison}
\boxed{\bigg(\partial_\Tt +\dfrac{c_s+v}{1+c_s v}\partial_\Tx \bigg)\phi =\dfrac{D}{\gamma^3 (1+c_s v)^3} \partial^2_\Tx \phi \, .}
\end{equation}
To model the sound wavepacket generated by the kick, we also need to boost the source term $\delta(t)\delta(x)$, and we obtain the retarded Green function
\begin{equation}\label{boostunzione}
\phi_{\text{Bob}}(\Tt,\Tx)=\Theta(\Tt) \sqrt{\dfrac{\gamma(1+c_s v)}{4\pi D\Tt}} e^{-\frac{\gamma^3 (1+c_s v)^3}{4D\Tt} \left(\Tx- \frac{c_s+v}{1+c_s v} \Tt\right)^2} \, .
\end{equation}
Expressed in Alice's coordinates, this becomes
\begin{equation}
\phi_{\text{Bob}}(t,x)=\Theta(t+vx) \sqrt{\dfrac{1+c_s v}{4\pi D(t+vx)}} e^{-\frac{1+c_s v}{4D(t+vx)} (x-c_s t)^2} \, .
\end{equation}
Again, no Lorentz factor appears, so $\phi_{\text{Bob}}(t,x)$ has a convergent limit as $v\rightarrow \pm 1$. The comparison between $\phi_{\text{Alice}}$, $\phi_{\text{Bob}}$, and $\phi_{\text{Burnett}}$ is provided in figure \ref{fig:soundiamo}. For Bob's speed, we consider both cases $v\rightarrow \pm 1$, since the problem is no longer symmetric under the 180$^\text{o}$ rotation $x\rightarrow -x$. For the transport parameters, we pick $c_s=1/\sqrt{3}$ and $\beta=\sqrt{2}$.

\subsection{Quantitative comparison}

Figure \ref{fig:soundiamo} shows that, for sound waves, the timescale for Alice to agree with Burnett is much longer than in pure diffusion. In fact, while in the case of diffusion there was good agreement already at $t/D=25$, this time we need to wait all the way till $t/D=500$ to reach the same level of accuracy. This is no surprise since, as we said earlier, the disagreement between Alice and Burnett now decays like $\sqrt{D/t}$ (compared to the $D/t$ decay law in pure diffusion).

Let us now comment on Bob's results. The main novelty is that now there are two different ``Bobs'', whose models perform very differently. The first Bob (red) travels with speed $1^-$ in a direction contrary to the wave, and his predictions are close to those of Alice already at $t/D=100$. The second Bob (magenta) ``outruns'' the sound wave, and his predictions perform rather poorly, maintaining visible disagreement with Alice and Burnett at $t/D=1000$. 

The above trends can be explained with a quick calculation. First, let us note that, in the case of Alice, Burnett, and any other higher-order theory of the form $\partial_t \phi=\sum_n c_n \partial_x^n \phi$, the baricenter of the wavepacket travels along a sound-type worldline, which emanates from the event where the kick was applied, i.e.
\begin{equation}
\dfrac{\int x\phi_{\text{Burnett}} \, dx}{\int \phi_{\text{Burnett}} \, dx}=\dfrac{\int x\phi_{\text{Alice}}  \, dx}{\int \phi_{\text{Alice}} \, dx} = c_s t \, . 
\end{equation}
This can be shown explicitly from equation \eqref{grinburnettwithsound}, invoking the well-known fact that multiplication by $x$ is dual to differentiation in Fourier space. On the other hand, in Bob's theory, the baricenter of the emergent wavepacket does \textit{not} travel on that same worldline. To understand why, consider Bob's solution in Bob's frame, i.e. equation \eqref{boostunzione}. Since its profile is even in $\Tilde{x}$, we know that the tails of the wavepacket are perfectly symmetric as viewed by Bob. However, when we boost back to the rest frame of the medium, one tail spreads earlier than the other tail, due to the relativity of simultaneity. As a result, more diffusion happens towards one side. The baricenter is thus displaced by a constant amount, which exists already at $t=0$, and which can be evaluated analytically:
\begin{equation}\label{displazione}
\dfrac{\int x\phi_{\text{Bob}}  \, dx}{\int \phi_{\text{Bob}} \, dx} =c_s t +\dfrac{2vD}{1+c_s v}  \, .    
\end{equation}
This offset should be contrasted with the size of the wavepacket at that same time $t$, which is $x{-}c_s t\sim \sqrt{Dt}$. Hence, the relative discrepancy due to the $v-$dependent displacement \eqref{displazione} is 
\begin{equation}
\dfrac{x_{\text{Bob}}-x_{\text{Alice}}}{\sqrt{Dt}}\sim  \dfrac{2v}{1+c_s v} \, \sqrt{\dfrac{D}{t}} \, .
\end{equation}
In the case where Bob moves contrary to sound ($v\rightarrow 1$), the above formula gives a 5\% disagreement for $t/D\sim 600$. In the case where he moves in the same sense as the wave ($v\rightarrow -1$), 5\% disagreement is reached for $t/D\sim 9000$.

\section{Conclusions}

We investigated how accurately the ``approximately boosted'' diffusion equation \eqref{DensityFrame} captures the underlying physics. Our analysis shows that, at relativistic velocities, this model performs notably worse than the standard diffusion equation \eqref{diffundo}. Such discrepancy arises from the structure of the truncation errors: transforming \eqref{diffundo} into \eqref{DensityFrame} introduces derivative terms that are of lower order than those discarded when deriving \eqref{diffundo} from the microscopic theory. Furthermore, we also verified that, from the perspective of the moving observer, the timescale for \eqref{DensityFrame} to become valid is Lorentz-dilated, and not contracted (contrary to the claims in \cite{Basar:2024qxd}).

\begin{figure}
    \centering
\includegraphics[width=0.45\linewidth]{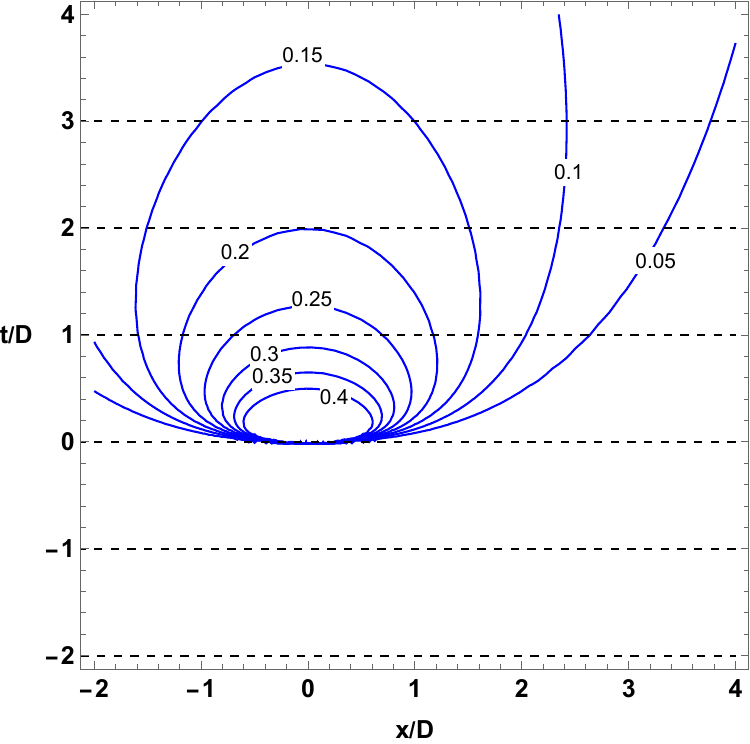}
\includegraphics[width=0.45\linewidth]{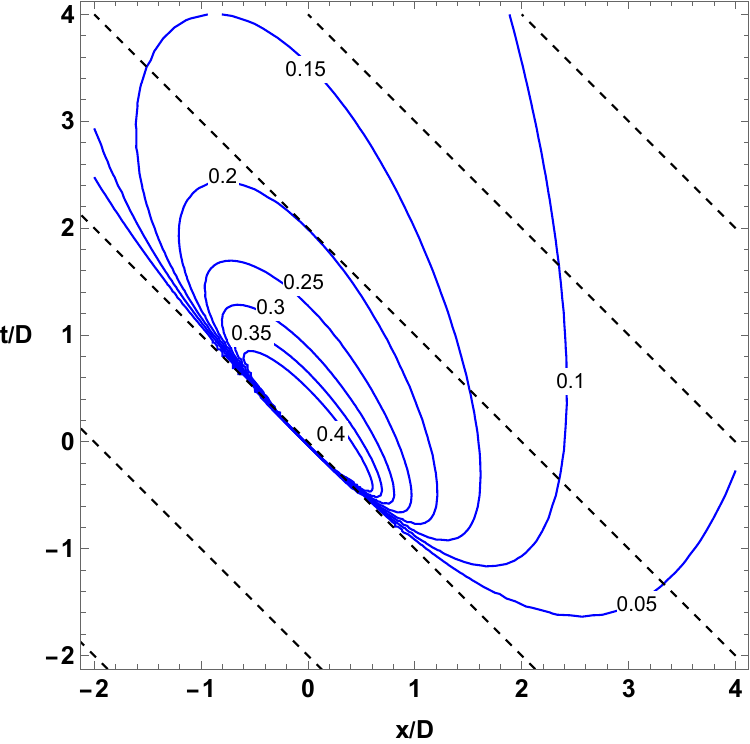}
\caption{Minkowski diagram illustrating how to construct the retarded Green function of Bob [who uses \eqref{DensityFrame}] starting from that of Alice [who uses \eqref{diffundo}], all plotted in Alice's frame. First, draw the contours of Alice's Green function (left panel, blue). These describe a superluminal diffusive process, whose tails spread along hyperplanes at constant Alice's time (dashed). Then, shift vertically every point of these contours by an amount equal to $-vx$, namely $(t,x)\rightarrow (t-vx,x)$. The result is Bob's Green function (right panel, blue), whose shape is ``tilted'', since the tails spread on hyperplanes at constant Bob's time (dashed). In the above example, we are working in the limit where Bob moves at speed $-1^+$.}
    \label{fig:contours}
\end{figure}

Despite the above limitations, equation \eqref{DensityFrame} performs significantly better than one might have expected. In fact, we found that there exist well-defined dynamical regimes (i.e. well-specified regions of spacetime in a given process) where \textit{all} observers who use \eqref{DensityFrame} arrive at mutually consistent outcomes. This happens because the truncation error in going from \eqref{diffundo} to \eqref{DensityFrame}, expressed in the rest frame of the medium, grows with powers of $v$, and not of $\gamma$. This is immediately apparent when we rewrite Bob's Green function in Alice's coordinates (see equation \eqref{Bobbonone}), where the \textit{only} source of disagreement between different observers is the relativity of simultaneity (see figure \ref{fig:contours} for a graph). The same can also be argued directly from the equation of motion. In fact, if we express \eqref{DensityFrame} in Alice's coordinates, we obtain \eqref{boninAlicetrame}, which admits a well-defined (i.e. convergent) limit as $v\rightarrow 1$.

Of course, the formal scope of the analysis carried out in this paper is rather limited, since we have restricted our discussion to (1+1)-dimensional problems in the linear regime. Nevertheless, we believe that the intuition we built from these exercises may still be useful in future (more general) conversations.

\newpage
We conclude this article with a brief summary of the advantages and limitations of Non-Covariant Parabolic Theories (NCPTs), such as \eqref{DensityFrame} and density-frame hydrodynamics \cite{Armas:2020mpr,Basar:2024qxd,Bhambure:2024gnf,Bhambure:2024axa}, compared to Covariant Hyperbolic Theories (CHTs) like Israel-Stewart~\cite{Israel_Stewart_1979} or BDNK~\cite{BemficaDNDefinitivo2020}. Based on our current knowledge, the following statements hold:
\begin{itemize}
\item In a given reference frame, NCPTs emerge as the late-time, long-wavelength limit of solutions of CHTs, by the ``relaxation effect'' discussed in \cite{nagy1994behavior,LindblomRelaxation1996}. This has been verified explicitly here. 
\item The formal applicability regimes of NCPTs and BDNK are essentially the same (first-order in gradients \cite{Kovtun2019,BemficaDNDefinitivo2020}), but the truncation error may be larger in the former if the fluid moves very fast relative to the observer. Yet, such error remains finite as $v\rightarrow 1$, and can be made arbitrarily small by sending the gradients to zero. This is the main result of the present work.
\item In many (but not all \cite{Heller2014,GavassinoUniversalityI2023,GavassinoBurgers2023}) real-world substances, the formal regime of applicability of Israel-Stewart hydrodynamics can be shown to be considerably larger than that of BDNK and NCPTs \cite{Denicol_Relaxation_2011,GavassinoGENERIC2022,GavassinoFarFromBulk2023}, provided that the additional transport coefficients are computed carefully \cite{WagnerGavassino2023jgq}. 
\item An advantage of NCPTs is that their mathematical structure is similar to that of non-relativistic Navier-Stokes, which makes it easier to import ideas (concerning e.g. shock profiles \cite{Catlfish1979,Catlfish1982} or turbulence \cite{Kolmogorov1991,Falkovich2006}) from the non-relativistic literature.
\item Another advantage is that the solutions of NCPTs are smoother and less singular than those of CHTs. This is due to their parabolic nature \cite[\S 7.1.3]{EvansPDEsBook}.
\item A disadvantage of NCPTs is that they can be solved only with initial data on hypersurfaces that are at constant time relative to the observer who solves them. It follows that different observers need different initial information.
\item It is formally possible to write down NCPTs in a curved spacetime \cite{Bhambure:2024gnf}. However, it is not yet clear whether a system that couples NCPTs with a \textit{dynamical} metric is free of pathologies.
\end{itemize}

\section*{Acknowledgements}
\vspace{-0.2cm}

This work is supported by a Vanderbilt's Seeding Success Grant. I thank Jorge Noronha and Derek Teaney for reading the manuscript and providing useful feedback. 

\newpage

\appendix

\section{Fourier transform of Bob's Green function in Alice's frame}\label{appendixAAA}

The Green function \eqref{Bobbonone} is a solution of \eqref{boninAlicetrame} with a localized source at $t=0$. It follows that, if we Fourier-transform \eqref{Bobbonone} in space, we necessarily obtain two different results at respectively positive and negative times, of the form
\begin{equation}\label{FourieriamoPhiBobbobe}
\phi_{\text{Bob}}(t,k)=
\begin{cases}
\phi^{t<0}_{\text{Gapless}}(k) e^{-i\omega_\text{Gapless}(k)t}+\phi^{t<0}_{\text{Gapped}}(k) e^{-i\omega_\text{Gapped}(k) t} & \text{if }t<0 \, ,\\
\\
\phi^{t>0}_{\text{Gapless}}(k) e^{-i\omega_\text{Gapless}(k)t}+\phi^{t>0}_{\text{Gapped}}(k) e^{-i\omega_\text{Gapped}(k)t} & \text{if }t>0 \, ,\\
\end{cases}
\end{equation}
where $\omega_\text{Gapless}$ and $\omega_\text{Gapped}$ are the two dispersion relations of \eqref{boninAlicetrame}, given by
\begin{equation}
\begin{split}
\omega_\text{Gapless}={}& \dfrac{i}{2Dv^2}-\dfrac{k}{v} -\dfrac{i}{2Dv^2}\sqrt{1+4iDvk} \, , \\
\omega_\text{Gapped}={}& \dfrac{i}{2Dv^2}-\dfrac{k}{v} +\dfrac{i}{2Dv^2}\sqrt{1+4iDvk} \, . \\
\end{split}
\end{equation}
On the other hand, a theorem by \citet{Hiscock_Insatibility_first_order} tells us that, since $\phi_\text{Bob}(t,x)$ does not blow up as $t\rightarrow +\infty$, the gapped contribution to $\phi_{\text{Bob}}(t,k)$ at positive times must vanish, i.e. $\phi^{t>0}_{\text{Gapped}}(k)=0$. Thus, the only dispersion relation that matters when we compare Alice's and Bob's predictions is $\omega_\text{Gapless}$, as argued in section \ref{vDkbeta}. 

To ensure the consistency of the above picture, we now proceed to compute the right-hand side of \eqref{FourieriamoPhiBobbobe} explicitly. We need to evaluate the following Fourier integral:
\begin{equation}
\phi_{\text{Bob}}(t,k)=\int_{\mathbb{R}} \Theta(t+vx) \dfrac{e^{\frac{-x^2}{4D(t+vx)}-ikx}}{\sqrt{4\pi D (t+vx)}} dx\, .
\end{equation}
Thanks to the change of variables $y=x+t/v$, such integral becomes
\begin{equation}
\begin{split}
\phi_{\text{Bob}}(t,k)={}& \int_0^{+\infty}  \dfrac{e^{\frac{-(y-t/v)^2}{4Dvy}-ik(y-t/v)}}{\sqrt{4\pi D vy}} dy \\
={}& \dfrac{e^{\left(\frac{1}{2Dv^2}+i\frac{k}{v}\right) t}}{\sqrt{4\pi D v}} \int_0^{+\infty}  \dfrac{e^{-\left(\frac{1}{4Dv}+ik\right)y-\frac{t^2}{4Dv^3 y}}}{\sqrt{y}} dy \\
={}& \dfrac{e^{\left(\frac{1}{2Dv^2}+i\frac{k}{v}\right) t}}{\sqrt{4\pi D v}} \sqrt{\dfrac{\pi}{\frac{1}{4Dv}+ik}} e^{-2\sqrt{\left(\frac{1}{4Dv}+ik\right)\frac{t^2}{4Dv^3 }}}\\
={}& \dfrac{e^{\left(\frac{1}{2Dv^2}+i\frac{k}{v}\right) t -\frac{|t|}{2Dv^2}\sqrt{1+4ikDv}}}{\sqrt{1+4iDkv} } \, . \\
\end{split}
\end{equation}
This can be rewritten as a function defined by cases:
\begin{equation}
\phi_{\text{Bob}}(t,k)=
\dfrac{1}{\sqrt{1{+}4iDkv}}
\begin{cases}
e^{-i\omega_{\text{Gapped}}(k) t} & \text{if }t<0 \, , \\
e^{-i\omega_{\text{Gapless}}(k) t} & \text{if }t>0 \, . \\
\end{cases}
\end{equation}
As can be seen, $\phi^{t>0}_{\text{Gapped}}(k)$ vanishes, meaning that only the gapless modes contribute to $\phi_\text{Bob}(t,x)$ for positive times.

\bibliography{Biblio}

\begin{thebibliography}{43}%
\makeatletter
\providecommand \@ifxundefined [1]{%
 \@ifx{#1\undefined}
}%
\providecommand \@ifnum [1]{%
 \ifnum #1\expandafter \@firstoftwo
 \else \expandafter \@secondoftwo
 \fi
}%
\providecommand \@ifx [1]{%
 \ifx #1\expandafter \@firstoftwo
 \else \expandafter \@secondoftwo
 \fi
}%
\providecommand \natexlab [1]{#1}%
\providecommand \enquote  [1]{``#1''}%
\providecommand \bibnamefont  [1]{#1}%
\providecommand \bibfnamefont [1]{#1}%
\providecommand \citenamefont [1]{#1}%
\providecommand \href@noop [0]{\@secondoftwo}%
\providecommand \href [0]{\begingroup \@sanitize@url \@href}%
\providecommand \@href[1]{\@@startlink{#1}\@@href}%
\providecommand \@@href[1]{\endgroup#1\@@endlink}%
\providecommand \@sanitize@url [0]{\catcode `\\12\catcode `\$12\catcode `\&12\catcode `\#12\catcode `\^12\catcode `\_12\catcode `\%12\relax}%
\providecommand \@@startlink[1]{}%
\providecommand \@@endlink[0]{}%
\providecommand \url  [0]{\begingroup\@sanitize@url \@url }%
\providecommand \@url [1]{\endgroup\@href {#1}{\urlprefix }}%
\providecommand \urlprefix  [0]{URL }%
\providecommand \Eprint [0]{\href }%
\providecommand \doibase [0]{http://dx.doi.org/}%
\providecommand \selectlanguage [0]{\@gobble}%
\providecommand \bibinfo  [0]{\@secondoftwo}%
\providecommand \bibfield  [0]{\@secondoftwo}%
\providecommand \translation [1]{[#1]}%
\providecommand \BibitemOpen [0]{}%
\providecommand \bibitemStop [0]{}%
\providecommand \bibitemNoStop [0]{.\EOS\space}%
\providecommand \EOS [0]{\spacefactor3000\relax}%
\providecommand \BibitemShut  [1]{\csname bibitem#1\endcsname}%
\let\auto@bib@innerbib\@empty
\bibitem [{\citenamefont {Hiscock}\ and\ \citenamefont {Lindblom}(1985)}]{Hiscock_Insatibility_first_order}%
  \BibitemOpen
  \bibfield  {author} {\bibinfo {author} {\bibfnamefont {W.}~\bibnamefont {Hiscock}}\ and\ \bibinfo {author} {\bibfnamefont {L.}~\bibnamefont {Lindblom}},\ }\href {\doibase 10.1103/PhysRevD.31.725} {\bibfield  {journal} {\bibinfo  {journal} {Physical review D: Particles and fields}\ }\textbf {\bibinfo {volume} {31}},\ \bibinfo {pages} {725} (\bibinfo {year} {1985})}\BibitemShut {NoStop}%
\bibitem [{\citenamefont {{Kost{\"a}dt}}\ and\ \citenamefont {{Liu}}(2000)}]{Kost2000}%
  \BibitemOpen
  \bibfield  {author} {\bibinfo {author} {\bibfnamefont {P.}~\bibnamefont {{Kost{\"a}dt}}}\ and\ \bibinfo {author} {\bibfnamefont {M.}~\bibnamefont {{Liu}}},\ }\href {\doibase 10.1103/PhysRevD.62.023003} {\bibfield  {journal} {\bibinfo  {journal} {\prd}\ }\textbf {\bibinfo {volume} {62}},\ \bibinfo {eid} {023003} (\bibinfo {year} {2000})},\ \Eprint {http://arxiv.org/abs/cond-mat/0010276} {arXiv:cond-mat/0010276 [cond-mat.stat-mech]} \BibitemShut {NoStop}%
\bibitem [{\citenamefont {Gavassino}\ \emph {et~al.}(2020)\citenamefont {Gavassino}, \citenamefont {Antonelli},\ and\ \citenamefont {Haskell}}]{GavassinoLyapunov_2020}%
  \BibitemOpen
  \bibfield  {author} {\bibinfo {author} {\bibfnamefont {L.}~\bibnamefont {Gavassino}}, \bibinfo {author} {\bibfnamefont {M.}~\bibnamefont {Antonelli}}, \ and\ \bibinfo {author} {\bibfnamefont {B.}~\bibnamefont {Haskell}},\ }\href {\doibase 10.1103/physrevd.102.043018} {\bibfield  {journal} {\bibinfo  {journal} {Physical Review D}\ }\textbf {\bibinfo {volume} {102}} (\bibinfo {year} {2020}),\ 10.1103/physrevd.102.043018}\BibitemShut {NoStop}%
\bibitem [{\citenamefont {Gavassino}(2023{\natexlab{a}})}]{GavassinoBounds2023}%
  \BibitemOpen
  \bibfield  {author} {\bibinfo {author} {\bibfnamefont {L.}~\bibnamefont {Gavassino}},\ }\href {\doibase 10.1016/j.physletb.2023.137854} {\bibfield  {journal} {\bibinfo  {journal} {Phys. Lett. B}\ }\textbf {\bibinfo {volume} {840}},\ \bibinfo {pages} {137854} (\bibinfo {year} {2023}{\natexlab{a}})},\ \Eprint {http://arxiv.org/abs/2301.06651} {arXiv:2301.06651 [hep-th]} \BibitemShut {NoStop}%
\bibitem [{\citenamefont {Cattaneo}(1958)}]{cattaneo1958}%
  \BibitemOpen
  \bibfield  {author} {\bibinfo {author} {\bibfnamefont {C.}~\bibnamefont {Cattaneo}},\ }\href {https://books.google.pl/books?id=mHGeQwAACAAJ} {\emph {\bibinfo {title} {Sur une forme de l'{\'e}quation de la chaleur {\'e}liminant le paradoxe d'une propagation instantan{\'e}e}}},\ Comptes rendus hebdomadaires des s{\'e}ances de l'Acad{\'e}mie des sciences\ (\bibinfo  {publisher} {Gauthier-Villars},\ \bibinfo {year} {1958})\BibitemShut {NoStop}%
\bibitem [{\citenamefont {Israel}\ and\ \citenamefont {Stewart}(1979)}]{Israel_Stewart_1979}%
  \BibitemOpen
  \bibfield  {author} {\bibinfo {author} {\bibfnamefont {W.}~\bibnamefont {Israel}}\ and\ \bibinfo {author} {\bibfnamefont {J.}~\bibnamefont {Stewart}},\ }\href {\doibase https://doi.org/10.1016/0003-4916(79)90130-1} {\bibfield  {journal} {\bibinfo  {journal} {Annals of Physics}\ }\textbf {\bibinfo {volume} {118}},\ \bibinfo {pages} {341 } (\bibinfo {year} {1979})}\BibitemShut {NoStop}%
\bibitem [{\citenamefont {{Liu}}\ \emph {et~al.}(1986)\citenamefont {{Liu}}, \citenamefont {{M{\"u}ller}},\ and\ \citenamefont {{Ruggeri}}}]{Liu1986}%
  \BibitemOpen
  \bibfield  {author} {\bibinfo {author} {\bibfnamefont {I.~S.}\ \bibnamefont {{Liu}}}, \bibinfo {author} {\bibfnamefont {I.}~\bibnamefont {{M{\"u}ller}}}, \ and\ \bibinfo {author} {\bibfnamefont {T.}~\bibnamefont {{Ruggeri}}},\ }\href {\doibase 10.1016/0003-4916(86)90164-8} {\bibfield  {journal} {\bibinfo  {journal} {Annals of Physics}\ }\textbf {\bibinfo {volume} {169}},\ \bibinfo {pages} {191} (\bibinfo {year} {1986})}\BibitemShut {NoStop}%
\bibitem [{\citenamefont {Geroch}\ and\ \citenamefont {Lindblom}(1991)}]{Geroch_Lindblom_1991_causal}%
  \BibitemOpen
  \bibfield  {author} {\bibinfo {author} {\bibfnamefont {R.}~\bibnamefont {Geroch}}\ and\ \bibinfo {author} {\bibfnamefont {L.}~\bibnamefont {Lindblom}},\ }\href {\doibase https://doi.org/10.1016/0003-4916(91)90063-E} {\bibfield  {journal} {\bibinfo  {journal} {Annals of Physics}\ }\textbf {\bibinfo {volume} {207}},\ \bibinfo {pages} {394 } (\bibinfo {year} {1991})}\BibitemShut {NoStop}%
\bibitem [{\citenamefont {{Baier}}\ \emph {et~al.}(2008)\citenamefont {{Baier}}, \citenamefont {{Romatschke}}, \citenamefont {{Thanh Son}}, \citenamefont {{Starinets}},\ and\ \citenamefont {{Stephanov}}}]{Baier2008}%
  \BibitemOpen
  \bibfield  {author} {\bibinfo {author} {\bibfnamefont {R.}~\bibnamefont {{Baier}}}, \bibinfo {author} {\bibfnamefont {P.}~\bibnamefont {{Romatschke}}}, \bibinfo {author} {\bibfnamefont {D.}~\bibnamefont {{Thanh Son}}}, \bibinfo {author} {\bibfnamefont {A.~O.}\ \bibnamefont {{Starinets}}}, \ and\ \bibinfo {author} {\bibfnamefont {M.~A.}\ \bibnamefont {{Stephanov}}},\ }\href {\doibase 10.1088/1126-6708/2008/04/100} {\bibfield  {journal} {\bibinfo  {journal} {Journal of High Energy Physics}\ }\textbf {\bibinfo {volume} {2008}},\ \bibinfo {eid} {100} (\bibinfo {year} {2008})},\ \Eprint {http://arxiv.org/abs/0712.2451} {arXiv:0712.2451 [hep-th]} \BibitemShut {NoStop}%
\bibitem [{\citenamefont {Denicol}\ \emph {et~al.}(2012)\citenamefont {Denicol}, \citenamefont {Niemi}, \citenamefont {Moln\'ar},\ and\ \citenamefont {Rischke}}]{Denicol2012Boltzmann}%
  \BibitemOpen
  \bibfield  {author} {\bibinfo {author} {\bibfnamefont {G.~S.}\ \bibnamefont {Denicol}}, \bibinfo {author} {\bibfnamefont {H.}~\bibnamefont {Niemi}}, \bibinfo {author} {\bibfnamefont {E.}~\bibnamefont {Moln\'ar}}, \ and\ \bibinfo {author} {\bibfnamefont {D.~H.}\ \bibnamefont {Rischke}},\ }\href {\doibase 10.1103/PhysRevD.85.114047} {\bibfield  {journal} {\bibinfo  {journal} {Phys. Rev. D}\ }\textbf {\bibinfo {volume} {85}},\ \bibinfo {pages} {114047} (\bibinfo {year} {2012})}\BibitemShut {NoStop}%
\bibitem [{\citenamefont {{Romatschke}}\ and\ \citenamefont {{Romatschke}}(2017)}]{Romatschke2017}%
  \BibitemOpen
  \bibfield  {author} {\bibinfo {author} {\bibfnamefont {P.}~\bibnamefont {{Romatschke}}}\ and\ \bibinfo {author} {\bibfnamefont {U.}~\bibnamefont {{Romatschke}}},\ }\href {\doibase 10.48550/arXiv.1712.05815} {\bibfield  {journal} {\bibinfo  {journal} {arXiv e-prints}\ ,\ \bibinfo {eid} {arXiv:1712.05815}} (\bibinfo {year} {2017})},\ \Eprint {http://arxiv.org/abs/1712.05815} {arXiv:1712.05815 [nucl-th]} \BibitemShut {NoStop}%
\bibitem [{\citenamefont {{Florkowski}}\ \emph {et~al.}(2018)\citenamefont {{Florkowski}}, \citenamefont {{Heller}},\ and\ \citenamefont {{Spali{\'n}ski}}}]{FlorkowskiReview2018}%
  \BibitemOpen
  \bibfield  {author} {\bibinfo {author} {\bibfnamefont {W.}~\bibnamefont {{Florkowski}}}, \bibinfo {author} {\bibfnamefont {M.~P.}\ \bibnamefont {{Heller}}}, \ and\ \bibinfo {author} {\bibfnamefont {M.}~\bibnamefont {{Spali{\'n}ski}}},\ }\href {\doibase 10.1088/1361-6633/aaa091} {\bibfield  {journal} {\bibinfo  {journal} {Reports on Progress in Physics}\ }\textbf {\bibinfo {volume} {81}},\ \bibinfo {eid} {046001} (\bibinfo {year} {2018})},\ \Eprint {http://arxiv.org/abs/1707.02282} {arXiv:1707.02282 [hep-ph]} \BibitemShut {NoStop}%
\bibitem [{\citenamefont {Gavassino}\ and\ \citenamefont {Antonelli}(2021)}]{GavassinoFronntiers2021}%
  \BibitemOpen
  \bibfield  {author} {\bibinfo {author} {\bibfnamefont {L.}~\bibnamefont {Gavassino}}\ and\ \bibinfo {author} {\bibfnamefont {M.}~\bibnamefont {Antonelli}},\ }\href {\doibase 10.3389/fspas.2021.686344} {\bibfield  {journal} {\bibinfo  {journal} {Front. Astron. Space Sci.}\ }\textbf {\bibinfo {volume} {8}},\ \bibinfo {pages} {686344} (\bibinfo {year} {2021})},\ \Eprint {http://arxiv.org/abs/2105.15184} {arXiv:2105.15184 [gr-qc]} \BibitemShut {NoStop}%
\bibitem [{\citenamefont {Gavassino}\ \emph {et~al.}(2024)\citenamefont {Gavassino}, \citenamefont {Disconzi},\ and\ \citenamefont {Noronha}}]{GavassinoUniversalityI2023}%
  \BibitemOpen
  \bibfield  {author} {\bibinfo {author} {\bibfnamefont {L.}~\bibnamefont {Gavassino}}, \bibinfo {author} {\bibfnamefont {M.~M.}\ \bibnamefont {Disconzi}}, \ and\ \bibinfo {author} {\bibfnamefont {J.}~\bibnamefont {Noronha}},\ }\href {\doibase 10.1103/PhysRevLett.132.222302} {\bibfield  {journal} {\bibinfo  {journal} {Phys. Rev. Lett.}\ }\textbf {\bibinfo {volume} {132}},\ \bibinfo {pages} {222302} (\bibinfo {year} {2024})},\ \Eprint {http://arxiv.org/abs/2302.03478} {arXiv:2302.03478 [nucl-th]} \BibitemShut {NoStop}%
\bibitem [{\citenamefont {Gavassino}(2022)}]{GavassinoSuperluminal2021}%
  \BibitemOpen
  \bibfield  {author} {\bibinfo {author} {\bibfnamefont {L.}~\bibnamefont {Gavassino}},\ }\href {\doibase 10.1103/PhysRevX.12.041001} {\bibfield  {journal} {\bibinfo  {journal} {Phys. Rev. X}\ }\textbf {\bibinfo {volume} {12}},\ \bibinfo {pages} {041001} (\bibinfo {year} {2022})}\BibitemShut {NoStop}%
\bibitem [{\citenamefont {Rauch}(1991)}]{Rauch_book}%
  \BibitemOpen
  \bibfield  {author} {\bibinfo {author} {\bibfnamefont {J.}~\bibnamefont {Rauch}},\ }\href {https://doi.org/10.1007/978-1-4612-0953-9} {\emph {\bibinfo {title} {{Partial Differential Equations}}}},\ Graduate Texts in Mathematics\ (\bibinfo  {publisher} {Springer},\ \bibinfo {address} {New York, NY},\ \bibinfo {year} {1991})\BibitemShut {NoStop}%
\bibitem [{\citenamefont {Wald}(1984)}]{Wald}%
  \BibitemOpen
  \bibfield  {author} {\bibinfo {author} {\bibfnamefont {R.~M.}\ \bibnamefont {Wald}},\ }\href {https://cds.cern.ch/record/106274} {\emph {\bibinfo {title} {{General relativity}}}}\ (\bibinfo  {publisher} {Chicago Univ. Press},\ \bibinfo {address} {Chicago, IL},\ \bibinfo {year} {1984})\BibitemShut {NoStop}%
\bibitem [{\citenamefont {Courant}\ and\ \citenamefont {Hilbert}(1989)}]{CourantHilbert2_book}%
  \BibitemOpen
  \bibfield  {author} {\bibinfo {author} {\bibfnamefont {R.}~\bibnamefont {Courant}}\ and\ \bibinfo {author} {\bibfnamefont {D.}~\bibnamefont {Hilbert}},\ }\href@noop {} {\emph {\bibinfo {title} {{Methods of Mathematical Physics, Vol 2: Partial Differential Equations}}}}\ (\bibinfo  {publisher} {John Wiley and Sons},\ \bibinfo {address} {New York, NY},\ \bibinfo {year} {1989})\BibitemShut {NoStop}%
\bibitem [{\citenamefont {Nagy}\ \emph {et~al.}(1994)\citenamefont {Nagy}, \citenamefont {Ortiz},\ and\ \citenamefont {Reula}}]{nagy1994behavior}%
  \BibitemOpen
  \bibfield  {author} {\bibinfo {author} {\bibfnamefont {G.~B.}\ \bibnamefont {Nagy}}, \bibinfo {author} {\bibfnamefont {O.~E.}\ \bibnamefont {Ortiz}}, \ and\ \bibinfo {author} {\bibfnamefont {O.~A.}\ \bibnamefont {Reula}},\ }\href@noop {} {\bibfield  {journal} {\bibinfo  {journal} {Journal of Mathematical Physics}\ }\textbf {\bibinfo {volume} {35}},\ \bibinfo {pages} {4334} (\bibinfo {year} {1994})}\BibitemShut {NoStop}%
\bibitem [{\citenamefont {Armas}\ and\ \citenamefont {Jain}(2021)}]{Armas:2020mpr}%
  \BibitemOpen
  \bibfield  {author} {\bibinfo {author} {\bibfnamefont {J.}~\bibnamefont {Armas}}\ and\ \bibinfo {author} {\bibfnamefont {A.}~\bibnamefont {Jain}},\ }\href {\doibase 10.21468/SciPostPhys.11.3.054} {\bibfield  {journal} {\bibinfo  {journal} {SciPost Phys.}\ }\textbf {\bibinfo {volume} {11}},\ \bibinfo {pages} {054} (\bibinfo {year} {2021})},\ \Eprint {http://arxiv.org/abs/2010.15782} {arXiv:2010.15782 [hep-th]} \BibitemShut {NoStop}%
\bibitem [{\citenamefont {Ba\c{s}ar}\ \emph {et~al.}(2024)\citenamefont {Ba\c{s}ar}, \citenamefont {Bhambure}, \citenamefont {Singh},\ and\ \citenamefont {Teaney}}]{Basar:2024qxd}%
  \BibitemOpen
  \bibfield  {author} {\bibinfo {author} {\bibfnamefont {G.}~\bibnamefont {Ba\c{s}ar}}, \bibinfo {author} {\bibfnamefont {J.}~\bibnamefont {Bhambure}}, \bibinfo {author} {\bibfnamefont {R.}~\bibnamefont {Singh}}, \ and\ \bibinfo {author} {\bibfnamefont {D.}~\bibnamefont {Teaney}},\ }\href {\doibase 10.1103/PhysRevC.110.044903} {\bibfield  {journal} {\bibinfo  {journal} {Phys. Rev. C}\ }\textbf {\bibinfo {volume} {110}},\ \bibinfo {pages} {044903} (\bibinfo {year} {2024})},\ \Eprint {http://arxiv.org/abs/2403.04185} {arXiv:2403.04185 [nucl-th]} \BibitemShut {NoStop}%
\bibitem [{\citenamefont {Bhambure}\ \emph {et~al.}(2024{\natexlab{a}})\citenamefont {Bhambure}, \citenamefont {Singh},\ and\ \citenamefont {Teaney}}]{Bhambure:2024gnf}%
  \BibitemOpen
  \bibfield  {author} {\bibinfo {author} {\bibfnamefont {J.}~\bibnamefont {Bhambure}}, \bibinfo {author} {\bibfnamefont {R.}~\bibnamefont {Singh}}, \ and\ \bibinfo {author} {\bibfnamefont {D.}~\bibnamefont {Teaney}},\ }\href@noop {} {\  (\bibinfo {year} {2024}{\natexlab{a}})},\ \Eprint {http://arxiv.org/abs/2412.10306} {arXiv:2412.10306 [nucl-th]} \BibitemShut {NoStop}%
\bibitem [{\citenamefont {Bhambure}\ \emph {et~al.}(2024{\natexlab{b}})\citenamefont {Bhambure}, \citenamefont {Mazeliauskas}, \citenamefont {Paquet}, \citenamefont {Singh}, \citenamefont {Singh}, \citenamefont {Teaney},\ and\ \citenamefont {Zhou}}]{Bhambure:2024axa}%
  \BibitemOpen
  \bibfield  {author} {\bibinfo {author} {\bibfnamefont {J.}~\bibnamefont {Bhambure}}, \bibinfo {author} {\bibfnamefont {A.}~\bibnamefont {Mazeliauskas}}, \bibinfo {author} {\bibfnamefont {J.-F.}\ \bibnamefont {Paquet}}, \bibinfo {author} {\bibfnamefont {R.}~\bibnamefont {Singh}}, \bibinfo {author} {\bibfnamefont {M.}~\bibnamefont {Singh}}, \bibinfo {author} {\bibfnamefont {D.}~\bibnamefont {Teaney}}, \ and\ \bibinfo {author} {\bibfnamefont {F.}~\bibnamefont {Zhou}},\ }\href@noop {} {\  (\bibinfo {year} {2024}{\natexlab{b}})},\ \Eprint {http://arxiv.org/abs/2412.10303} {arXiv:2412.10303 [nucl-th]} \BibitemShut {NoStop}%
\bibitem [{\citenamefont {{Geroch}}(1995)}]{Geroch1995}%
  \BibitemOpen
  \bibfield  {author} {\bibinfo {author} {\bibfnamefont {R.}~\bibnamefont {{Geroch}}},\ }\href {\doibase 10.1063/1.530958} {\bibfield  {journal} {\bibinfo  {journal} {Journal of Mathematical Physics}\ }\textbf {\bibinfo {volume} {36}},\ \bibinfo {pages} {4226} (\bibinfo {year} {1995})}\BibitemShut {NoStop}%
\bibitem [{\citenamefont {{Lindblom}}(1996)}]{LindblomRelaxation1996}%
  \BibitemOpen
  \bibfield  {author} {\bibinfo {author} {\bibfnamefont {L.}~\bibnamefont {{Lindblom}}},\ }\href {\doibase 10.1006/aphy.1996.0036} {\bibfield  {journal} {\bibinfo  {journal} {Annals of Physics}\ }\textbf {\bibinfo {volume} {247}},\ \bibinfo {pages} {1} (\bibinfo {year} {1996})},\ \Eprint {http://arxiv.org/abs/gr-qc/9508058} {arXiv:gr-qc/9508058 [gr-qc]} \BibitemShut {NoStop}%
\bibitem [{\citenamefont {McLennan}(1973)}]{MCLennanBurnett1973}%
  \BibitemOpen
  \bibfield  {author} {\bibinfo {author} {\bibfnamefont {J.~A.}\ \bibnamefont {McLennan}},\ }\href {\doibase 10.1103/PhysRevA.8.1479} {\bibfield  {journal} {\bibinfo  {journal} {Phys. Rev. A}\ }\textbf {\bibinfo {volume} {8}},\ \bibinfo {pages} {1479} (\bibinfo {year} {1973})}\BibitemShut {NoStop}%
\bibitem [{\citenamefont {Hartman}\ \emph {et~al.}(2017)\citenamefont {Hartman}, \citenamefont {Hartnoll},\ and\ \citenamefont {Mahajan}}]{Hartman:2017hhp}%
  \BibitemOpen
  \bibfield  {author} {\bibinfo {author} {\bibfnamefont {T.}~\bibnamefont {Hartman}}, \bibinfo {author} {\bibfnamefont {S.~A.}\ \bibnamefont {Hartnoll}}, \ and\ \bibinfo {author} {\bibfnamefont {R.}~\bibnamefont {Mahajan}},\ }\href {\doibase 10.1103/PhysRevLett.119.141601} {\bibfield  {journal} {\bibinfo  {journal} {Phys. Rev. Lett.}\ }\textbf {\bibinfo {volume} {119}},\ \bibinfo {pages} {141601} (\bibinfo {year} {2017})},\ \Eprint {http://arxiv.org/abs/1706.00019} {arXiv:1706.00019 [hep-th]} \BibitemShut {NoStop}%
\bibitem [{\citenamefont {{Heller}}\ \emph {et~al.}(2023)\citenamefont {{Heller}}, \citenamefont {{Serantes}}, \citenamefont {{Spali{\'n}ski}},\ and\ \citenamefont {{Withers}}}]{HellerBounds2023}%
  \BibitemOpen
  \bibfield  {author} {\bibinfo {author} {\bibfnamefont {M.~P.}\ \bibnamefont {{Heller}}}, \bibinfo {author} {\bibfnamefont {A.}~\bibnamefont {{Serantes}}}, \bibinfo {author} {\bibfnamefont {M.}~\bibnamefont {{Spali{\'n}ski}}}, \ and\ \bibinfo {author} {\bibfnamefont {B.}~\bibnamefont {{Withers}}},\ }\href {\doibase 10.1103/PhysRevLett.130.261601} {\bibfield  {journal} {\bibinfo  {journal} {\prl}\ }\textbf {\bibinfo {volume} {130}},\ \bibinfo {eid} {261601} (\bibinfo {year} {2023})},\ \Eprint {http://arxiv.org/abs/2212.07434} {arXiv:2212.07434 [hep-th]} \BibitemShut {NoStop}%
\bibitem [{\citenamefont {Gavassino}(2024)}]{GavassinoChapmanEnskog2024xwf}%
  \BibitemOpen
  \bibfield  {author} {\bibinfo {author} {\bibfnamefont {L.}~\bibnamefont {Gavassino}},\ }\href {\doibase 10.1103/PhysRevD.110.094012} {\bibfield  {journal} {\bibinfo  {journal} {Phys. Rev. D}\ }\textbf {\bibinfo {volume} {110}},\ \bibinfo {pages} {094012} (\bibinfo {year} {2024})}\BibitemShut {NoStop}%
\bibitem [{\citenamefont {Struchtrup}\ and\ \citenamefont {Taheri}(2011)}]{Struchtrup2011}%
  \BibitemOpen
  \bibfield  {author} {\bibinfo {author} {\bibfnamefont {H.}~\bibnamefont {Struchtrup}}\ and\ \bibinfo {author} {\bibfnamefont {P.}~\bibnamefont {Taheri}},\ }\href {\doibase 10.1093/imamat/hxr004} {\bibfield  {journal} {\bibinfo  {journal} {IMA Journal of Applied Mathematics}\ }\textbf {\bibinfo {volume} {76}},\ \bibinfo {pages} {672} (\bibinfo {year} {2011})},\ \Eprint {http://arxiv.org/abs/https://academic.oup.com/imamat/article-pdf/76/5/672/1835580/hxr004.pdf} {https://academic.oup.com/imamat/article-pdf/76/5/672/1835580/hxr004.pdf} \BibitemShut {NoStop}%
\bibitem [{\citenamefont {Bemfica}\ \emph {et~al.}(2022)\citenamefont {Bemfica}, \citenamefont {Disconzi},\ and\ \citenamefont {Noronha}}]{BemficaDNDefinitivo2020}%
  \BibitemOpen
  \bibfield  {author} {\bibinfo {author} {\bibfnamefont {F.~S.}\ \bibnamefont {Bemfica}}, \bibinfo {author} {\bibfnamefont {M.~M.}\ \bibnamefont {Disconzi}}, \ and\ \bibinfo {author} {\bibfnamefont {J.}~\bibnamefont {Noronha}},\ }\href {\doibase 10.1103/PhysRevX.12.021044} {\bibfield  {journal} {\bibinfo  {journal} {Phys. Rev. X}\ }\textbf {\bibinfo {volume} {12}},\ \bibinfo {pages} {021044} (\bibinfo {year} {2022})}\BibitemShut {NoStop}%
\bibitem [{\citenamefont {{Kovtun}}(2019)}]{Kovtun2019}%
  \BibitemOpen
  \bibfield  {author} {\bibinfo {author} {\bibfnamefont {P.}~\bibnamefont {{Kovtun}}},\ }\href {\doibase 10.1007/JHEP10(2019)034} {\bibfield  {journal} {\bibinfo  {journal} {Journal of High Energy Physics}\ }\textbf {\bibinfo {volume} {2019}},\ \bibinfo {eid} {34} (\bibinfo {year} {2019})},\ \Eprint {http://arxiv.org/abs/1907.08191} {arXiv:1907.08191 [hep-th]} \BibitemShut {NoStop}%
\bibitem [{\citenamefont {Heller}\ \emph {et~al.}(2014)\citenamefont {Heller}, \citenamefont {Janik}, \citenamefont {Spali\ifmmode~\acute{n}\else \'{n}\fi{}ski},\ and\ \citenamefont {Witaszczyk}}]{Heller2014}%
  \BibitemOpen
  \bibfield  {author} {\bibinfo {author} {\bibfnamefont {M.~P.}\ \bibnamefont {Heller}}, \bibinfo {author} {\bibfnamefont {R.~A.}\ \bibnamefont {Janik}}, \bibinfo {author} {\bibfnamefont {M.}~\bibnamefont {Spali\ifmmode~\acute{n}\else \'{n}\fi{}ski}}, \ and\ \bibinfo {author} {\bibfnamefont {P.}~\bibnamefont {Witaszczyk}},\ }\href {\doibase 10.1103/PhysRevLett.113.261601} {\bibfield  {journal} {\bibinfo  {journal} {Phys. Rev. Lett.}\ }\textbf {\bibinfo {volume} {113}},\ \bibinfo {pages} {261601} (\bibinfo {year} {2014})}\BibitemShut {NoStop}%
\bibitem [{\citenamefont {Gavassino}(2023{\natexlab{b}})}]{GavassinoBurgers2023}%
  \BibitemOpen
  \bibfield  {author} {\bibinfo {author} {\bibfnamefont {L.}~\bibnamefont {Gavassino}},\ }\href {\doibase 10.1088/1361-6382/ace587} {\bibfield  {journal} {\bibinfo  {journal} {Class. Quant. Grav.}\ }\textbf {\bibinfo {volume} {40}},\ \bibinfo {pages} {165008} (\bibinfo {year} {2023}{\natexlab{b}})},\ \Eprint {http://arxiv.org/abs/2304.05455} {arXiv:2304.05455 [nucl-th]} \BibitemShut {NoStop}%
\bibitem [{\citenamefont {Denicol}\ \emph {et~al.}(2011)\citenamefont {Denicol}, \citenamefont {Noronha}, \citenamefont {Niemi},\ and\ \citenamefont {Rischke}}]{Denicol_Relaxation_2011}%
  \BibitemOpen
  \bibfield  {author} {\bibinfo {author} {\bibfnamefont {G.~S.}\ \bibnamefont {Denicol}}, \bibinfo {author} {\bibfnamefont {J.}~\bibnamefont {Noronha}}, \bibinfo {author} {\bibfnamefont {H.}~\bibnamefont {Niemi}}, \ and\ \bibinfo {author} {\bibfnamefont {D.~H.}\ \bibnamefont {Rischke}},\ }\href {\doibase 10.1103/PhysRevD.83.074019} {\bibfield  {journal} {\bibinfo  {journal} {Phys. Rev. D}\ }\textbf {\bibinfo {volume} {83}},\ \bibinfo {pages} {074019} (\bibinfo {year} {2011})}\BibitemShut {NoStop}%
\bibitem [{\citenamefont {{Gavassino}}\ and\ \citenamefont {{Antonelli}}(2023)}]{GavassinoGENERIC2022}%
  \BibitemOpen
  \bibfield  {author} {\bibinfo {author} {\bibfnamefont {L.}~\bibnamefont {{Gavassino}}}\ and\ \bibinfo {author} {\bibfnamefont {M.}~\bibnamefont {{Antonelli}}},\ }\href {\doibase 10.1088/1361-6382/acc165} {\bibfield  {journal} {\bibinfo  {journal} {Classical and Quantum Gravity}\ }\textbf {\bibinfo {volume} {40}},\ \bibinfo {eid} {075012} (\bibinfo {year} {2023})},\ \Eprint {http://arxiv.org/abs/2209.12865} {arXiv:2209.12865 [gr-qc]} \BibitemShut {NoStop}%
\bibitem [{\citenamefont {Gavassino}\ and\ \citenamefont {Noronha}(2024)}]{GavassinoFarFromBulk2023}%
  \BibitemOpen
  \bibfield  {author} {\bibinfo {author} {\bibfnamefont {L.}~\bibnamefont {Gavassino}}\ and\ \bibinfo {author} {\bibfnamefont {J.}~\bibnamefont {Noronha}},\ }\href {\doibase 10.1103/PhysRevD.109.096040} {\bibfield  {journal} {\bibinfo  {journal} {Phys. Rev. D}\ }\textbf {\bibinfo {volume} {109}},\ \bibinfo {pages} {096040} (\bibinfo {year} {2024})},\ \Eprint {http://arxiv.org/abs/2305.04119} {arXiv:2305.04119 [gr-qc]} \BibitemShut {NoStop}%
\bibitem [{\citenamefont {Wagner}\ and\ \citenamefont {Gavassino}(2024)}]{WagnerGavassino2023jgq}%
  \BibitemOpen
  \bibfield  {author} {\bibinfo {author} {\bibfnamefont {D.}~\bibnamefont {Wagner}}\ and\ \bibinfo {author} {\bibfnamefont {L.}~\bibnamefont {Gavassino}},\ }\href {\doibase 10.1103/PhysRevD.109.016019} {\bibfield  {journal} {\bibinfo  {journal} {Phys. Rev. D}\ }\textbf {\bibinfo {volume} {109}},\ \bibinfo {pages} {016019} (\bibinfo {year} {2024})},\ \Eprint {http://arxiv.org/abs/2309.14828} {arXiv:2309.14828 [nucl-th]} \BibitemShut {NoStop}%
\bibitem [{\citenamefont {{Caflisch}}(1979)}]{Catlfish1979}%
  \BibitemOpen
  \bibfield  {author} {\bibinfo {author} {\bibfnamefont {R.~E.}\ \bibnamefont {{Caflisch}}},\ }\href@noop {} {\bibfield  {journal} {\bibinfo  {journal} {Communications in Pure Applied Mathematics}\ }\textbf {\bibinfo {volume} {32}},\ \bibinfo {pages} {521} (\bibinfo {year} {1979})}\BibitemShut {NoStop}%
\bibitem [{\citenamefont {{Caflisch}}\ and\ \citenamefont {{Nicolaenko}}(1982)}]{Catlfish1982}%
  \BibitemOpen
  \bibfield  {author} {\bibinfo {author} {\bibfnamefont {R.~E.}\ \bibnamefont {{Caflisch}}}\ and\ \bibinfo {author} {\bibfnamefont {B.}~\bibnamefont {{Nicolaenko}}},\ }\href {\doibase 10.1007/BF01206009} {\bibfield  {journal} {\bibinfo  {journal} {Communications in Mathematical Physics}\ }\textbf {\bibinfo {volume} {86}},\ \bibinfo {pages} {161} (\bibinfo {year} {1982})}\BibitemShut {NoStop}%
\bibitem [{\citenamefont {{Kolmogorov}}(1991)}]{Kolmogorov1991}%
  \BibitemOpen
  \bibfield  {author} {\bibinfo {author} {\bibfnamefont {A.~N.}\ \bibnamefont {{Kolmogorov}}},\ }\href {\doibase 10.1098/rspa.1991.0075} {\bibfield  {journal} {\bibinfo  {journal} {Proceedings of the Royal Society of London Series A}\ }\textbf {\bibinfo {volume} {434}},\ \bibinfo {pages} {9} (\bibinfo {year} {1991})}\BibitemShut {NoStop}%
\bibitem [{\citenamefont {{Falkovich}}\ and\ \citenamefont {{Sreenivasan}}(2006)}]{Falkovich2006}%
  \BibitemOpen
  \bibfield  {author} {\bibinfo {author} {\bibfnamefont {G.}~\bibnamefont {{Falkovich}}}\ and\ \bibinfo {author} {\bibfnamefont {K.~R.}\ \bibnamefont {{Sreenivasan}}},\ }\href {\doibase 10.1063/1.2207037} {\bibfield  {journal} {\bibinfo  {journal} {Physics Today}\ }\textbf {\bibinfo {volume} {59}},\ \bibinfo {pages} {43} (\bibinfo {year} {2006})}\BibitemShut {NoStop}%
\bibitem [{\citenamefont {Evans}(1997)}]{EvansPDEsBook}%
  \BibitemOpen
  \bibfield  {author} {\bibinfo {author} {\bibfnamefont {L.~C.}\ \bibnamefont {Evans}},\ }\href@noop {} {\emph {\bibinfo {title} {Partial Differential Equations}}}\ (\bibinfo  {publisher} {American Mathematical Society},\ \bibinfo {address} {Berkeley},\ \bibinfo {year} {1997})\BibitemShut {NoStop}%
\end{thebibliography}%

\end{document}